\documentclass{emulateapj}
\usepackage{float}
\usepackage{amsmath}
\usepackage{amssymb}
\usepackage{latexsym}
\usepackage{textcomp}
\usepackage{graphicx}
\usepackage{subfigure}
\usepackage{epsfig}
\usepackage{natbib}

\newcommand{\ha}{\textrm{H}\ensuremath{\alpha}}
\newcommand{\hb}{\textrm{H}\ensuremath{\beta}}

\newcommand{\hei}{[\textrm{He}\,\textsc{i}]}
\newcommand{\nii}{[\textrm{N}~\textsc{ii}]}

\newcommand{\oii}{[\textrm{O}\,\textsc{ii}]}
\newcommand{\oiii}{[\textrm{O}~\textsc{iii}]}
\newcommand{\hii}{\textrm{H}~\textsc{ii}}
\newcommand{\sii}{[\textrm{S}~\textsc{ii}]}
\newcommand{\oi}{[\textrm{O}~\textsc{i}]}

\newcommand{\oiiilam}{[\textrm{O}~\textsc{iii}]~\ensuremath{\lambda5007}}
\newcommand{\niilam}{[\textrm{N}~\textsc{ii}]~\ensuremath{\lambda6584}} 
\newcommand{\niilamboth}{[\textrm{N}\,\textsc{ii}]\,\ensuremath{\lambda\lambda6548,6584}} 
\newcommand{\oiilam}{[\textrm{O}~\textsc{ii}]~\ensuremath{\lambda3727}}
\newcommand{\oiilamboth}{[\textrm{O}\,\textsc{ii}]\,\ensuremath{\lambda\lambda3726,3729}}
\newcommand{\oilam}{[\textrm{O}~\textsc{i}]~\ensuremath{\lambda6300}}
\newcommand{\siilam}{[\textrm{S}~\textsc{ii}]~\ensuremath{\lambda\lambda6717,6731}}
\newcommand{\neiiilam}{[\textrm{Ne}~\textsc{iii}]~\ensuremath{\lambda3869}} 
\newcommand{\nevlam}{[\textrm{Ne}~\textsc{v}]~\ensuremath{\lambda3525}} 
\newcommand{\oiiihb}{[\textrm{O}~\textsc{iii}]/\textrm{H}\ensuremath{\beta}}

\newcommand{\oiiilamboth}{[\textrm{O}\,\textsc{iii}]\,\ensuremath{\lambda\lambda4959,5007}}

\newcommand{\ergs}{\textrm{erg\,s$^{-1}$}}

\newcommand{\Mstar}{$M_{\star}$}
\newcommand{\Msun}{$M_{\sun}$}

\slugcomment{ApJ; Accepted for publication}

\shorttitle{High-z BPT and MEx Diagrams}
\shortauthors{Juneau et al.}

\begin{document}

\title{AGN Emission Line Diagnostics and the Mass-Metallicity Relation up to Redshift $\lowercase{z}\sim2$: \\ the Impact of Selection Effects and Evolution}

\author{\sc St\'{e}phanie Juneau\altaffilmark{1,2}}
\altaffiltext{1}{CEA-Saclay, DSM/IRFU/SAp, 91191 Gif-sur-Yvette, France}
\altaffiltext{2}{stephanie.juneau@cea.fr}

\author{\sc Fr\'ed\'eric Bournaud\altaffilmark{1}}

\author{\sc St\'ephane Charlot\altaffilmark{3}}
\altaffiltext{3}{UPMC-CNRS, UMR 7095, Institut d'Astrophysique de Paris, 75014, Paris, France}

\author{\sc Emanuele Daddi\altaffilmark{1}}

\author{\sc David Elbaz\altaffilmark{1}}

\author{\sc Jonathan R. Trump\altaffilmark{4,5}}
\altaffiltext{4}{University of California Observatories/Lick Observatory, University of California, Santa Cruz, CA 95064, USA}
\altaffiltext{5}{Hubble Fellow; Department of Astronomy and Astrophysics, 525 Davey Lab, The Pennsylvania State University, University Park, PA 16802, USA}

\author{\sc Jarle Brinchmann\altaffilmark{6}}
\altaffiltext{6}{Leiden Observatory, Leiden University, P.O. Box 9513, 2300 RA Leiden, The Netherlands}

\author{\sc Mark Dickinson\altaffilmark{7}}
\altaffiltext{7}{National Optical Astronomy Observatory, 950 North Cherry Avenue, Tucson, AZ 85719, USA}

\author{\sc Pierre-Alain Duc\altaffilmark{1}}

\author{\sc Raphael Gobat\altaffilmark{1}}

\author{\sc Ingrid Jean-Baptiste\altaffilmark{1,8}}

\author{\sc \'Emeric Le Floc'h\altaffilmark{1}}

\author{\sc M. D. Lehnert\altaffilmark{8,3}}
\altaffiltext{8}{GEPI, Observatoire de Paris, UMR 8111, CNRS, Universit\'e Paris Diderot, 5 place Jules Janssen, 92190, Meudon, France}

\author{\sc Camilla Pacifici\altaffilmark{9,3}}
\altaffiltext{9}{Yonsei University Observatory, Yonsei University, Seoul 120-749, Republic of Korea}

\author{\sc Maurilio Pannella\altaffilmark{1}}

\author{\sc Corentin Schreiber\altaffilmark{1}}

\begin{abstract}

Emission line diagnostic diagrams probing the ionization sources in galaxies, such as the 
Baldwin-Phillips-Terlevich (BPT) diagram, have been used extensively to distinguish AGN from 
purely star-forming galaxies.  Yet, they remain poorly understood at higher redshifts.  
We shed light on this issue with an empirical approach based on a $z\sim0$ reference sample 
built from $\sim$300,000 SDSS galaxies, from which we mimic 
selection effects due to typical emission line detection limits at higher redshift.  
We combine this low-redshift reference sample with a simple prescription for luminosity evolution of 
the global galaxy population to predict the loci of high-redshift galaxies on the BPT and 
Mass-Excitation (MEx) diagnostic diagrams. The predicted bivariate distributions agree remarkably 
well with direct observations of galaxies out to $z\sim1.5$, including the 
observed stellar mass-metallicity ($MZ$) relation evolution.  
As a result, we infer that high-redshift star-forming galaxies are consistent with having {\it normal} ISM 
properties out to $z\sim1.5$, after accounting for selection effects and line luminosity evolution.  
Namely, their optical line ratios and gas-phase metallicities are comparable to that of low-redshift galaxies with equivalent emission-line 
luminosities. In contrast, AGN narrow-line regions may show a shift toward lower metallicities at higher redshift. 
While a physical evolution of the ISM conditions is not ruled out for purely star-forming galaxies, and 
may be more important starting at $z\gtrsim2$, we find that reliably quantifying this evolution 
is hindered by selections effects.  The recipes provided here may serve as a basis for future studies 
toward this goal. Code to predict the loci of 
galaxies on the BPT and MEx diagnostic diagrams, and the $MZ$ relation as a function of emission 
line luminosity limits, is made publicly available.

\end{abstract}

\keywords{galaxies: active --- galaxies: evolution --- galaxies: fundamental parameters --- 
          galaxies: Seyfert --- galaxies: star formation}

\section{Introduction}

Nebular emission lines can reveal crucial information on the ionized gas content in  
galaxies.  In particular, several optical emission line diagnostics have been 
developed to probe gas properties such as metallicity, ionization parameter, electron 
density and temperature \citep{ost06}, 
which can in turn provide additional insights on the source of ionization of the gas. 
An important application is thus the identification of active galactic nuclei (AGNs), 
which leave strong signatures on nebular line ratios such as \oiiilam/\hb\ and/or 
\niilam/\ha.  These two line ratios form the most traditional version of the 
BPT diagram \citep{bpt,vei87}.  The latter has been calibrated with both a theoretical 
approach \citep{kew01,sta06,kew13a} and empirically with low-redshift galaxies \citep{kau03c}.  

There are now questions about the applicability of low-redshift nebular 
line diagnostics to higher-redshift objects.  
A number of studies suggest that high-redshift galaxies are offset from the locus 
of low-redshift reference samples on the standard BPT diagram 
(\oiiilam/\hb\ vs. \niilam/\ha) \citep[e.g.,][]{sha05,erb06,tru13,new14,hol14}. 
While there are a few hypotheses, the cause of this offset is not yet fully explained.  
For example, it was suggested that high-redshift galaxies may have had different \hii\ 
region conditions (such as electron densities, temperatures, pressures, etc.) relative 
to the bulk of star-forming galaxies \citep{bri08,liu08,hai09,leh09,rig11,ly13}.
It was suggested \citep{leh13,shi13} that this may be due to galaxies globally 
forming their stars with a higher surface density in the past, which has been inferred 
from infrared luminosity surface densities \citep{red12} and galaxy infrared SED 
fitting \citep{mag12}.

However, other studies claim that the offsets on excitation diagrams are 
instead caused by an increased contribution from AGN \citep{gro06,wri10,tru11}, 
which would shift the galaxies in a similar way \citep{kew13a}.  
If there were a higher incidence of AGN in galaxies 
in the past we may expect a steeper ionization profile and thus varying 
emission line strengths.  It is crucial to disentangle the source 
of ionization in galaxies (young stars vs. AGN) in order to interpret and derive 
important quantities in galaxy evolution studies like star formation rates (SFRs), 
metallicities, and gas dynamics, but also to understand the interplay between 
black hole growth and stellar growth in galaxies.

Furthermore, another complication arises because intermediate- and high-redshift galaxy samples 
used thus far may suffer from strong selection biases due to the emission line detection limits. 
Relative to existing large spectroscopic sample at low redshifts (e.g., SDSS), only galaxies 
with intrinsically luminous lines can be detected at intermediate to high redshifts. 
These potential selection biases have been mostly neglected thus far, and will be explored in this 
Paper along with genuine evolutionary trends.  As we will show, emission line detection limits add 
complexity to the problem, but not taking them into account can yield misguided interpretations of 
how galaxy properties evolve with redshift. 

In addition to the traditional BPT-\nii\ diagnostic diagram, we revisit an alternative diagram 
using stellar mass in place of \nii/\ha\ \citep[the Mass-Excitation (MEx) 
diagnostic diagram from][hereafter J11]{jun11}.  The MEx diagram has the advantage of requiring 
only the \oiii/\hb\ emission lines, which are more widely separated in wavelength and therefore 
easier to resolve spectroscopically than \ha\ and \nii.  Furthermore, they be can observed to higher 
redshift in any given wavelength regime.  In optical spectra, \nii/\ha\ are available out to $z\sim0.45$ 
whereas \oiii/\hb\ can be observed out to $z\sim0.9$.  Similarly, NIR spectra in the 
$K$ band cover \nii/\ha\ out to $z\sim2.5$ but \oiii/\hb\ out to $z\sim3.7$.  
Another advantage of the MEx diagram is its probabilistic approach.  For a given location on the MEx 
plane, and given the measurement errors, the MEx diagram yields the probability that the 
galaxy hosts an AGN.  This method has a built-in uncertainty in the sense that ambiguous cases 
will have a low or intermediate AGN probability, and is well suited for statistical studies because 
the AGN probabilities can be used as statistical weigths to weigh for (or against) AGN.  
On the other hand, one might expect the MEx diagram to be more sensitive to evolution of the 
stellar-mass metallicity ($MZ$) relation \citep{sav05,sha05,erb06,yab12,zah13}, than the traditional BPT.  
We will show that an improved treatment of emission-line detection limit mitigates such bias by directly 
accounting for appropriate gas-phase metallicities when building a tailored low-redshift 
comparison sample for each survey.

The aim of this Paper is twofold.  First, we provide improved AGN diagnostics that account for 
redshift-dependent effects.  More specifically, we revisit both the original BPT diagram and 
more recent MEx diagram in order to disentangle selection and evolution effects, 
and to improve their applicability to a broad range of redshifts.  In addition, this work  
reveals insight into the ISM conditions in higher redshift galaxies, once the selection effects are 
taken into account.

The Paper is organized as follows.  
We describe the galaxy samples used for low-redshift calibration and higher redshift applications 
in Section~\ref{sec:sample}, followed by the low-redshift revision of the MEx demarcations 
in Section~\ref{sec:revised}.  The results (Section~\ref{sec:result}) include empirical predictions 
of the redshift evolution of the BPT and MEx diagrams including both genuine evolution and selection 
effects due to line detection limits (Section~\ref{sec:detlim}).  These predictions are confronted 
with observations out to $z\sim2$ (Section~\ref{sec:hiz}), and compared to theoretical predictions 
from \citet{kew13a} in Section~\ref{sec:theo}. 
The implications for the high-redshift application of emission line diagnostic diagrams are 
discussed in Section~\ref{sec:discu}, including the stellar mass-metallicity relation, 
before the main findings are summarized in Section~\ref{summ}.  Throughout this paper, we assume a 
flat $\Lambda$CDM cosmology  ($\Omega_m = 0.3$, $\Omega_{\Lambda} = 0.7$, and $h = 0.7$) and a 
\citet{cha03} initial mass function (IMF).

\section{Sample Selection}\label{sec:sample}

\subsection{Low-redshift galaxy sample}\label{sec:lozsample}

The low-redshift emission-line galaxy sample is built from the Sloan Digital Sky Survey (SDSS) DR7 \citep{aba09}.
The first selection criteria ensure that the galaxies are primary targets ($SCIENCEPRIMARY = 1$), 
and have a redshift determination in the range of interest ($0.04<z<0.2$). 
The lower and higher redshift cuts are imposed in order to, respectively, avoid strong aperture 
effects ($z>0.04$) and offer a good compromise between detecting galaxies with intrinsically weak 
lines and obtaining better statistics on Seyferts ($z<0.2$), following a similar approach 
to that of \citet{kew06,yua10}. 

From a primary sample of 426,367 galaxies, we further select 299,098 galaxies for which the 
\oiiilam/\hb\ and \niilam/\ha\ line ratios are detected.  Removing 5,469 galaxies with a missing 
or invalid stellar mass ($<10^6~M_{\sun}$), we obtain a sample of 293,629 galaxies with a median 
redshift of $z=0.09$. We call this emission-line galaxy sample the {\it $z\sim0$ SDSS prior sample} 
because it will be used as a set of priors to calculate the probability of galaxies hosting AGN 
given certain observables (namely stellar mass and \oiii/\hb\ ratio) following the MEx method 
developed by J11. 

Emission line fluxes were obtained from the Value Added Catalogs developed by the 
Max-Planck Institute for Astronomy (Garching) and 
John Hopkins University (MPA/JHU)\footnote{http://www.mpa-garching.mpg.de/SDSS/DR7/},  
following the methodology described by \citet{tre04}.  We apply two corrections to the 
measurements listed in the Value Added Catalogs.  First, we apply a correction to \hb\ fluxes 
which were found to be underestimated by $\Delta {\rm EW} = 0.35$\AA \citep{gro12} because of 
the change in stellar population models between DR4 and DR7 \citep[Charlot \& Bruzual 2007 (CB07), 
instead of][(BC03)]{bru03}.  
The corrected \hb\ line fluxes are thus consistent with the use of the BC03 models for fitting the 
stellar continuum (and stellar absorption), like was done with SDSS DR4 spectra.  
Second, we augment the catalog's formal line flux uncertainties to represent more closely the true 
uncertainties by comparing emission line measurements made on duplicate observations of the same 
galaxies (Appendix~\ref{app:err}).

Traditionally, emission-line galaxies are selected to have detections with a given signal-to-noise 
ratio (S/N), such as S/N$>3$, for all four BPT-\nii\ lines (\oiiilam, \hb, \niilam, \ha).  
To obtain a more complete sample, and because the BPT diagram deals with line ratios, the detection 
limit is applied to the emission line ratios rather than to the individual lines. 
We require that the emission line {\it ratios} have S/N$>2.12$ ($=3/\sqrt{2}$), 
where the lower limit is equivalent to each line being detected at exactly $3\sigma$.  This cut 
furthermore includes combinations of a poorly-detected line ($<3\sigma$) with a strongly-detected 
line provided the overall ratio is constrained to better than S/N$=$2.12.  This modified approach 
yields a $\sim20$\% larger and therefore more complete census of emission-line galaxies, spanning 
a wider range of intrinsic properties. Relative to the more traditional approach, 
we include more numerous massive, metal-rich star-forming galaxies, as well as LINERs and 
\emph{retired} galaxies, all of which tend to have comparatively faint \oiiilam\ lines \citep{cid10,cid11}.

This $z\sim0$ SDSS prior sample is used to perform a new base calibration of the MEx diagnostic 
diagram at low redshift. 
However, when comparing to higher redshift samples, we revert to the typical method of imposing an 
individual line S/N$>3$ cut for \oiii\ and \ha.  This choice 
does not have a noticeable impact on the results presented in this paper, which concerns 
galaxies with fairly luminous lines that tend to be individually detected above $3\sigma$.  
For example, at log($L_{\ha,\oiii}$[\ergs])$>39.9$, the lowest luminosity cut employed in 
this paper, only 0.8\% of emission-line galaxies fail the S/N$>$3 criterion for either \ha\ or 
\oiii.

Lastly, the emission-line sample is classified for the presence of AGN using the BPT-\nii\ diagnostic diagram \citep{bpt}.  
Galaxies below and to the left of the \citet{kau03c} demarcation are considered star-forming, 
while the galaxies above and to the right of the \citet{kew01} demarcation are AGN, and the 
galaxies between the lines are often called composites. Galaxies from the \citet{kew01} AGN region can be further 
classified into Seyfert 2 (Sy2) or LINER based on their location on the BPT-\sii\ diagram \citep{vei87} 
using, for example, the division developed by \citet{kew06}\footnote{All four classes (Star-Forming, 
Composite, LINER, and Sy2) are distinguished in the MEx classification software distributed on the world wide 
web in order to allow users flexibility in the choice of populations that they wish to consider or 
remove from their samples.  IDL (Interactive Data Language) code and instructions are available 
here: https://sites.google.com/site/agndiagnostics/home/mex}.  
In this work, however, the LINER population is naturally removed by the emission line limits, effectively 
excluding the weak emission lines of LINER galaxies.  Thus, most of our analyses apply to the star-forming, 
composite, and Sy2 classes.  

We choose to treat composite and Sy2 galaxies as both hosting AGN.  As argued by \citet{sal07}, 
galaxies in the composite region have higher SFRs compared to those in the AGN region, such that the bulk of 
the difference from AGN to composite may simply be a weaker contrast between star formation and AGN lines.  
There is also good X-ray evidence (for both individual sources and stacked data) that most composite galaxies 
genuinely host AGNs \citep[e.g.,][]{jun11,tro11,tru11}.  A few authors argue that some composite galaxies have 
line ratios influenced by shock activity rather than AGN \citep{ric11,new14}, but these tend to be rare starbursts 
and not galaxies representative of the bulk population at $z\sim0$ or $z\sim2$.  Instead, most composite 
galaxies are likely to correspond to a transition phase between starburst- and AGN-dominated systems \citep{yua10}, 
in agreement with the concept of varying contrast between AGN and star formation emission and our 
categorization of composite galaxies as AGN.

\subsection{High-redshift galaxy samples}\label{sec:hizsamples}

The intermediate to high redshift galaxies used in this work were selected from the following:
\begin{itemize}
\item{$0.3<z<1$ galaxies
with $R_{AB}<24.3$ and $<24.1$ from the TKRS and DEEP2 redshift surveys, respectively (J11);}
\item{$z\sim1.4$ galaxies with $K<23.9$, $1.2<z_{phot}<1.6$, \Mstar$>10^{9.5}$~\Msun 
from the SXDS/UDS\footnote{Subaru XMM Deep Survey/UKIDSS Ultra Deep Survey} fields 
 with NIR spectra \citep[][hereafter Y12]{yab12};}
\item{$z\sim1.5$ emission-line selected galaxies from the GOODS-S field 
 with NIR spectra \citep[][hereafter T13]{tru13};}
\item{$z\sim2$ galaxies from the SINS/zC-SINF survey, GOODS-N and Q2343 fields, with NIR spectra, 
 and selected by \citet[][hereafter N14]{new14}.}
\end{itemize}

At $0.3<z<1$, optical spectroscopy was obtained from the Team Keck Redshift 
Survey\footnote{http://tkserver.keck.hawaii.edu/tksurvey/} 
\citep[TKRS][]{wir04} in GOODS-N, and from the DEEP2 Galaxy Redshift Survey 
\citep[hereafter DEEP2;][]{dav03,new12} in the Extended Growth Strip (EGS).
Both spectroscopic surveys were obtained with the Keck/DEIMOS spectrograph 
\citep{fab03} and the data were reduced with the same pipeline \citep{coo12}.  

Ancillary observations came from the Great Observatories Origins Deep 
Survey\footnote{http://www.stsci.edu/science/goods/} \citep[GOODS,][]{gia04}, and the 
All-wavelength Extended Groth strip International Survey \citep[AEGIS,][]{dav07}\footnote{http://aegis.ucolick.org/}.    
As a reminder, stellar masses were calculated with UV-to-NIR SED fitting following the 
method described by \citet{sal07} when the photometry was available.
Otherwise, stellar masses were estimated from the rest-frame $K$-band luminosity 
as described by J11 in their Appendix~B. 
In GOODS-N, J11 used the following photometry: $UBVRIz$ taken from \citet{cap04} 
and $JK$ obtained with the Flamingos camera on the Mayall 4~m NOAO telescope.  
For galaxies in EGS, they used FUV, NUV (GALEX), $ugriz$ (CFHTLS), and $K$ (Palomar) 
\citep[see][]{sal09,gwy08,gwy11,bun06}. 
For EGS galaxies outside of the Canada-France-Hawaii Telescope Legacy Survey (CFHTLS) 
field-of-view, they used CFHT 12k $BRI$ photometry \citep{coi04}.

Intermediate redshift galaxies are split into two redshift bins, $0.3<z<0.6$ and $0.6<z<1$, 
and are used only with the MEx diagnostic diagram because the redder lines \nii\ and \ha\ are outside 
the observed range of those optical spectra at approximately $z>0.4$. The parent galaxy sample and data 
are described by J11. For this work, galaxies were further selected to have 
S/N$>3$ emission line fluxes for \oiii, while \hb\ can be either a $3\sigma$ upper limit 
or a $>3\sigma$ detection. As described by J11, \hb\ fluxes were corrected for Balmer absorption 
using BC03 models to subtract the continuum when the spectra had a sufficient S/N per pixel ($>3$) 
and otherwise using the median value of $2.8 (\pm0.9)$~\AA.  This correction changes the 
\oiii/\hb\ ratios by 0.08~dex (r.m.s).

\renewcommand{\thefootnote}{\alph{footnote}}

\begin{deluxetable*}{lrccl}
\tabletypesize{\scriptsize}
\tablecolumns{5}
\tablewidth{0pc}
\tablecaption{Definition of Galaxy Samples\label{tab:samples}}
\tablehead{
   \colhead{Sample}  &  \colhead{Number}  & \colhead{Redshift}  & \colhead{Flux Limit} &  \colhead{Comments} \\
   \colhead{ }  &  \colhead{ }  & \colhead{ }  & \colhead{\ergs\,${\rm cm^{-2}}$} &  \colhead{ } 
}
\startdata
$\langle z\rangle = 0.45$   &   1729    &  $0.3<z<0.6$     &  $2\times10^{-17}$ & GOODS-N and EGS fields (J11)  \\
$\langle z\rangle = 0.7$    &   1662    &  $0.6<z<1.0$     &  $2\times10^{-17}$ & GOODS-N and EGS fields (J11)  \\
$\langle z\rangle = 1.4$    &     32    &  $z\sim1.4$      &  $4\times10^{-17}$ & SXDS/UDS field (Y12) \\
$\langle z\rangle = 1.5$    &     36    &  $z\sim1.5$      &  $3\times10^{-17}$ & GOODS-S field (T13)  \\
$\langle z\rangle = 2$      &     22    &  $z\sim1.5-2.5$  &  $4\times10^{-17}$\tablenotemark{a} & SINS/zC-SINF, GOODS-N and Q2343 fields (N14)
\enddata
\tablenotetext{a}{The formal flux detection limit is not given by N14 but was estimated from the flux calibrated spectra 
shown in their article.}
\tablenotetext{}{Columns: (1) Sample name; (2) Number of galaxies; (3) Redshift range; (4) $3\sigma$ Flux detection limit for emission lines; (5) Comments}
\end{deluxetable*}

\renewcommand{\thefootnote}{\arabic{footnote}}

At $z>1$, near-infrared spectra are required to measure rest-frame optical 
lines used for both the MEx and full BPT diagrams.  
Those were observed with Subaru/FMOS in SXDX/UDS (Y12), with 
a combination of HST/WFC3 and Keck/MOSFIRE in GOODS-S (T13), and 
with VLT/SINFONI or LBT/LUCI1 in the $z\sim2$ sample (N14).

Ancillary data that were used in the work of Y12, T13 and N14 come from the Subaru XMM Deep Survey 
\citep[SXDS,][]{fur08} and UKIDSS Ultra Deep Survey \citep[UDS,][]{law07} for the 
$z\sim1.4$ sample, from the GOODS and CANDELS \citep{grog11,koe11} surveys for the $z\sim1.5$ sample, 
and from a more heterogeneous set of observations for the $z\sim2$ sample.  Detailed description of the latter 
can be found in the original SINS/zC-SINF survey descriptions \citep{for09,manc11} and in  
Sections~2.1 and 2.2 of N14.  

The stellar masses and emission line ratios were taken from the published work introducing 
the samples (J11, T13, N14) or from private communication (K. Yabe, 2012).  
Information on each galaxy sample is summarized in Table~\ref{tab:samples}.
All three $z>1$ galaxy samples were not corrected for stellar absorption.  
Y12 argue that it is negligible based on stellar population fitting around \ha\ 
(median correction of 4.2~\AA\ is small compared to their typical \ha\ equivalent widths of 200~\AA)
and based on previous observations. \citet{zah11} found a median value of 0.9~\AA\ for \hb\ 
from a stack of DEEP2 spectra at $0.75<z<0.82$, corresponding to a median shift downward by $\sim$0.03~dex.
While the line ratios published by N14 do not include Balmer absorption, they have estimated that \oiii/\hb\ 
would shift downward by $\sim$0.02 to 0.21~dex (mean of 0.08~dex) from their best fit SED-derived 
star-formation histories and ages.

Whenever available, multi-wavelength 
AGN classifications allow us to identify candidate AGNs independently from the emission line 
diagrams.  These include X-rays in most cases (J11, T13, N14), as well as IRAC colors \citep{ste05} 
and radio excess emission \citep{del13} for the $0.3<z<1$ galaxies, following the procedure described by 
\citet{jun13}. 
AGN identification by N14 also relied on other criteria including elevated emission line ratios 
in the central region of their spatially resolved emission line maps, or UV or mid-IR 
signatures \citep[][F{\"o}rster Schreiber et al., in prep.]{for09}. 
On their side, Y12 excluded X-ray sources \citep{ued08} from their sample.  
The main characteristics of each galaxy sample are listed in Table~\ref{tab:samples}.

\section{Revised MEx Diagnostic Diagram}\label{sec:revised}

The MEx diagnostic diagram is revisited following two modifications relative to the initial design.  First, 
the prior calibration sample is now built from SDSS DR7 instead of DR4.  Second, the emission line 
signal-to-noise criterion is applied to the line ratios rather than to the individual lines (Section~\ref{sec:lozsample}).  
Otherwise, the approach is very similar: we empirically determine dividing lines that follow transitional 
values of P(AGN), the probability of hosting an AGN according to the prior sample classified with the 
traditional BPT diagrams.  

\begin{figure}[h]
\epsscale{1.05} \plotone{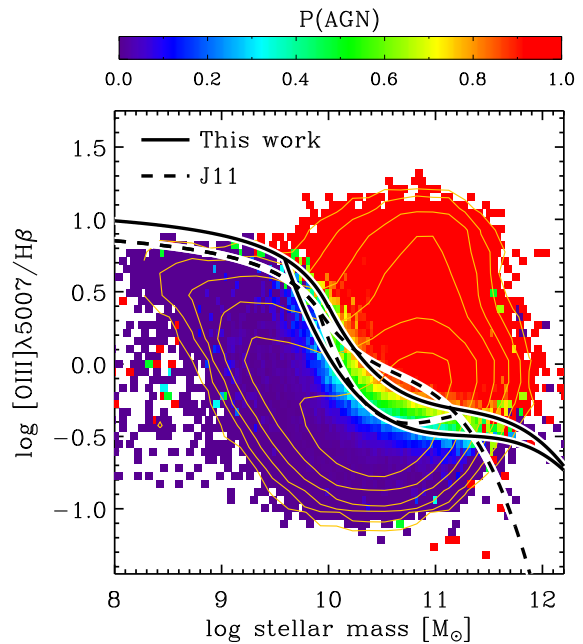}
\caption{
   MEx diagnostic diagram for the SDSS DR7 emission-line sample at $0.04<z<0.2$.  
   The contours mark the number density of galaxies in bins of 0.15\,dex$\times$0.15\,dex, 
   with the outermost contour corresponding to 10 galaxies per bin and a logarithm spacing 
   in steps of 0.5\,dex. The color scheme indicates the fraction of galaxies classified as 
   AGN using the BPT-\nii\ and \sii\ diagnostics, from 0.0 (purple) to 
   1.0 (red).  The demarcation lines are shown for the current sample (solid lines) as well 
   as for the original MEx diagram built from SDSS DR4 by J11 (dashed lines).
}\label{fig:newmex}
\end{figure}

At low stellar masses, the demarcation follows the upper envelope of the left-hand 
star-forming branch: $y = 0.375/(x - 10.4) + 1.14$, where $y \equiv$ log$(\oiiilam/\hb)$ and $x \equiv$ log$(M_{\star})$. 
This relation is fixed but the value of the transition mass ($M_{transi}$), where the low-mass relation connects 
with the high-mass end, is left as a free parameter of the fit. 
At high stellar masses, a third order polynomial is adjusted to pass through regions with $0.6 < P(AGN) < 0.85$ 
for the upper MEx curve, and with $0.3 < P(AGN) < 0.5$ for the lower MEx curve\footnote{The ranges of $P(AGN)$ values were chosen to have a similar number of pixels to perform the fitting on the MEx plane for the upper and lower curve, and to bracket the intermediate region characterized by a steep gradient from $P(AGN)\sim0.3$ to $P(AGN)\sim0.85$; see color bar on Figure~\ref{fig:newmex}.}.  
Using the IDL package \emph{mpfit} \citep{mar09} to solve for the best-fitting values, we obtain the following results.

The revised upper demarcation is defined as:
\begin{equation}\label{eq:up}
y = \left\{ \begin{array}{ll}
      0.375/(x - 10.5) + 1.14 & \mbox{if x\ensuremath{\leq}10} \\
      a_0 + a_1 x + a_2 x^2 + a_3 x^3 & \mbox{otherwise,}
      \end{array}
     \right. 
\end{equation}
where $y \equiv$ log$(\oiiilam/\hb)$ and $x \equiv$ log$(M_{\star})$. 
The coefficients are the following: $\{a_0, a_1, a_2, a_3\} = 
\{410.24, -109.333, 9.71731, -0.288244\}$.
Similarly, the lower curve is given by the following:
\begin{equation}\label{eq:lo}
y = \left\{ \begin{array}{ll}
      0.375/(x - 10.5) + 1.14 & \mbox{if x\ensuremath{\leq}9.6} \\
      a_0 + a_1 x + a_2 x^2 + a_3 x^3 & \mbox{otherwise.}
      \end{array}
     \right. 
\end{equation}
The coefficients are: $\{a_0, a_1, a_2, a_3\} = 
\{352.066, -93.8249, 8.32651, -0.246416\}$.

We show the updated demarcations along with the original ones in Figure~\ref{fig:newmex}. 
The demarcations are mostly used for vizualisation purposes and the number of AGNs is instead 
calculated based on the underlying bivariate distribution of the prior galaxy sample.  In this case, 
the $z\sim0$ SDSS DR7 emission-line sample (as defined in Section~\ref{sec:lozsample}) is shown.  
When calculating a number -- or fraction -- of AGNs in a given galaxy sample, the use of 
AGN probabilities is more appriopriate than the strict use of the demarcation lines, and can 
yield to a smaller number of AGNs.  This comes naturally from summing values of P(AGN)$\leq1$, 
but it should be kept in mind as it represents a difference from traditional use of AGN/SF diagnostic 
diagrams.

\section{Results}\label{sec:result}

\subsection{Effect of Emission-Line Detection Limit}\label{sec:detlim}

Applications of the BPT-\nii\ diagram at higher redshift must rely on NIR spectra with limited sensitivity \citep[e.g.][]{sha05,liu08,tru13}. 
In previous and current NIR spectroscopy studies, \ha\ and \oiii\ are the most commonly detected lines, 
while \nii\ and/or \hb\ are frequently undetected, yielding respectively to upper limits on \nii/\ha\ 
or lower limits on \oiii/\hb\ ratios.  
In what follows, we mimic such selection effects that arise within higher-redshift samples by requiring both \ha\ and \oiii\ in the $z\sim0$ SDSS sample to be more luminous than the emission line detection limits of intermediate to high-redshift optical and NIR spectroscopic surveys.  There is no constraint applied to \nii\ and \hb\ to allow for cases that would 
have an upper limit only for either or both of those two lines.

In the low-redshift SDSS sample, the \oiii\ line tends to be less luminous than \ha\ in the 
majority (97\%) of galaxies.  This means that requiring both \oiii\ and \ha\ to be 
more luminous than a common threshold is effectively an \oiii\ selection at the 97\% level. 
However, this trend only holds for 60-70\% of galaxies for current samples of higher redshift galaxies 
\citep[e.g.,][]{ly07,col13}.  For consistency across all redshifts considered in this work, we 
apply the line luminosity cut to both lines in the main part of this article, 
but we also consider various selections based on single emission 
lines or alternative line luminosity evolution in Appendix~\ref{app:altern}.
Some of the other scenarios yield similar results as there are more than one ways to 
select a comparison sample with better ressemblance to high-redshift galaxy surveys than using 
the full low-redshift SDSS survey as a comparison sample.  The latter results in a poor comparison, 
and should therefore be avoided in many studies. We release versatile code allowing the user to apply 
different scenarios tailored to surveys probing rest-frame optical lines (\hb\ and/or \oiii\ and/or \ha).

\begin{figure*}
\epsscale{0.9} \plotone{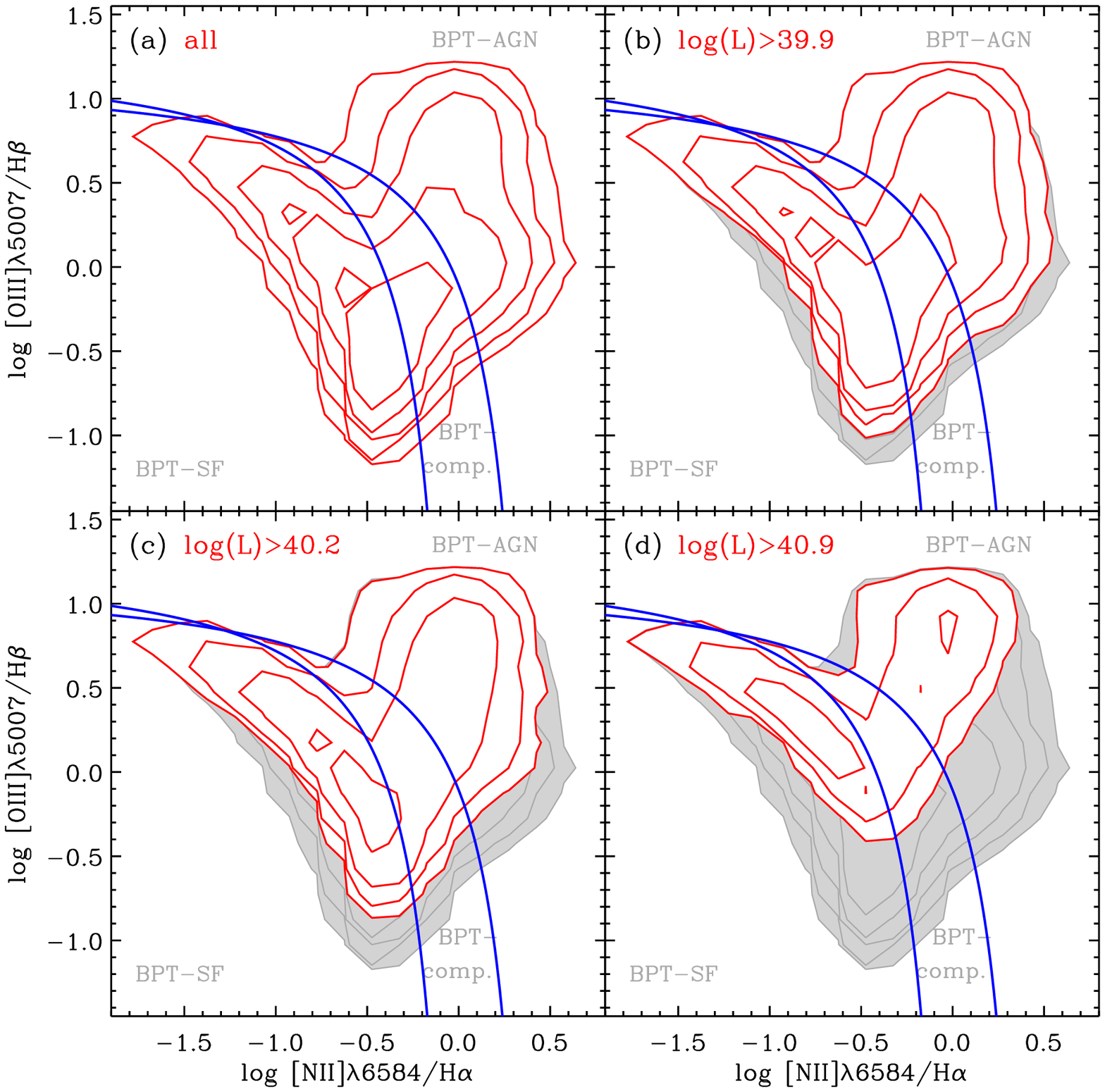}
\caption{
   BPT-\nii\ diagnostic diagram. The lower dividing line defines the
   upper envelope of star-forming galaxies \citep{kau03c} while the upper line
   from \citet{kew01} separates the most extreme AGN (above) from composite galaxies with
   a relatively higher SFR (between the lines).  Red contours show the 
   $z\sim0$ SDSS sample with 
   increasingly stricter luminosity thresholds applied to the \ha\ and
   \oiii\ lines: (a) no restriction, (b) log($L_{line}$[\ergs])$>39.9$
   (c) log($L_{line}$[\ergs])$>40.2$, (d) log($L_{line}$[\ergs])$>40.9$.  The underlying gray shaded
   contours show the full bivariate distribution identical to that in panel (a), 
   and correspond to the number density per bin (0.15\,dex$\times$0.15\,dex) with 
   logarithmic spacing (0.5\,dex).  The outermost contour shows 10 galaxies per bin.
}\label{fig:bpt}
\end{figure*}

\begin{figure*}
\epsscale{0.9} \plotone{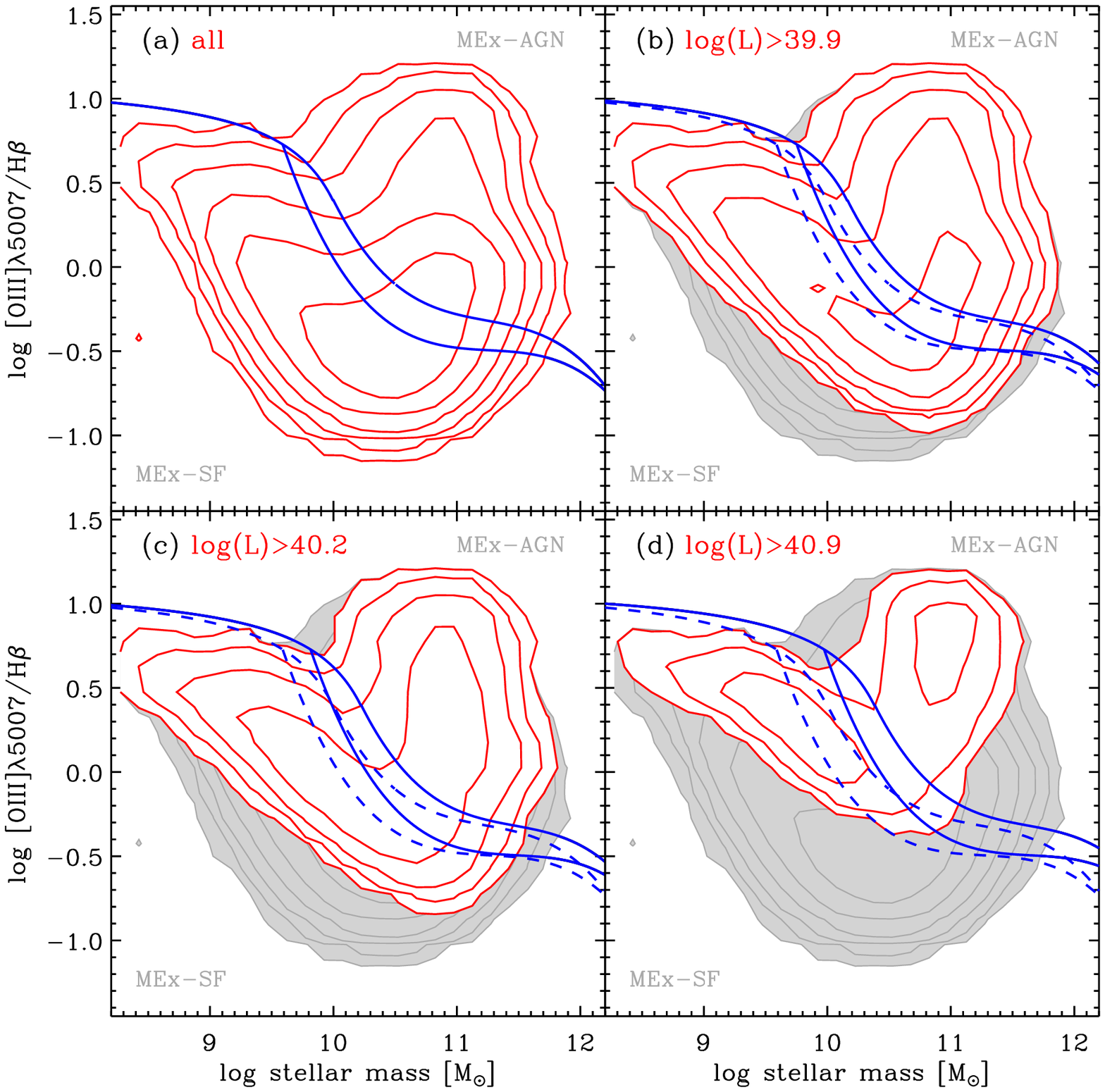}
\caption{
   MEx diagnostic diagram. 
   The dividing lines indicate regions corresponding to
   star-forming galaxies (below), MEx-intermediate galaxies (between)
   and MEx-AGN (above).  The demarcation lines derived for the full sample 
   are shown in panel (a) and in dashed lines in all other panels for reference. 
   As in Figure~\ref{fig:bpt}, red contours show
   the distributions of $z\sim0$ SDSS for varying line luminosity
   threshold (as labeled), while the gray shaded contours include the
   full $z\sim0$ prior sample.  Solid blue lines show the AGN/SF demarcations shifted 
   following the AGN fractions calculated from the BPT classification 
   (Appendix~\ref{sec:offset}). 
}\label{fig:mex}
\end{figure*}


The first test consists of applying increasingly brighter luminosity cuts to \ha\ and \oiii\ for their inclusion in the sample.  The results are respectively shown for the BPT-\nii\ and MEx diagrams in Figures~\ref{fig:bpt} and \ref{fig:mex}.  As the emission line detection luminosity threshold increases, the bivariate distributions shift upward on both the BPT-\nii\ and MEx diagrams. In other words, there is a bias against galaxies with low \oiii/\hb\ ratio or with high \nii/\ha\ when the latter is due to particularly weak \ha\ as occurs in LINERs and retired galaxies \citep{bri04,cid11}. 
The BPT demarcation lines appear to properly split the branches of the bivariate distributions in most cases.  
In details, the star-forming branch gets slightly closer to the Kauffmann line at the highest luminosity 
probed, suggesting a potential small shift toward higher \nii/\ha\ and/or higher \oiii/\hb, but this 
effect remains small ($<$0.2dex).  However, the MEx demarcation lines no longer trace the morphology of the branches when a high luminosity threshold is applied. Instead, the location of the split between the two branches appears to shift toward higher stellar masses as the luminosity threshold increases (Figure~\ref{fig:mex}). This shift arises because the MEx AGN probabilities 
(i.e., fraction of galaxies classified as BPT-AGN) depend on the emission line luminosities.  Thus, we recalibrate the 
MEx diagram to take this dependence into account in the AGN probability calculations.  We also calculate the corresponding 
offsets for the demarcation lines in Appendix~\ref{sec:offset} before reporting them on Figure~\ref{fig:mex}.

\subsection{Emission-Line Diagnostics at Higher Redshift}\label{sec:hiz}

At moderate or high redshifts, galaxy samples may be subject to both emission line detection 
limits and evolution of the emission-line galaxy population as a whole.  The former can be 
taken into account by applying the equivalent detection limit to the prior sample as described 
in Sections~\ref{sec:detlim}. If the low-redshift galaxy sample includes identical galaxies, 
applying the detection threshold should reproduce the properties of the higher redshift galaxies.  
However, this approach may fail if the low- and high-redshift samples are intrinsically different 
due to, e.g., significant evolution of the bulk of the galaxy population.  

One form of evolution can be probed through emission line luminosity functions, which show 
a general fading of the emission-line galaxy population with cosmic time \citep[e.g.][]{sob13,col13}.  
In this work, we adopt an evolution of the characteristic luminosity 
given by $\log(L^*_{\ha}(z)) =  \log(L^*_{z=0} \times (1+z)^{2.27})$, obtained by fitting to a compilation 
of $L^*$ values from the literature (Appendix~\ref{app:altern}).  We assume that the fading of 
the global galaxy population is traced by the fading of $L^*$ and that by accounting for it, 
we select galaxies that have comparable line luminosities relative to the mean of the evolving population. 
In practice, this means that we compare high-redshift galaxies with lower redshift galaxies 
that have slightly less luminous emission lines.

We determine an effective minimum luminosity threshold to define the prior sample by using 
$L_{threshold} = L_{detection} - \Delta(L^*)$, where $\Delta(L^*)$ is the difference between $L^*$ 
of the high-redshift sample and that of the comparison prior sample ($z=0.09$).  
This fading may be explained by decreasing normalization of the 
$M_{\star}-$SFR main sequence with cosmic time \citep{noe07,elb07,dad07,elb11}, itself a consequence 
of a decreasing gas fraction in galaxies \citep[e.g.,][]{sar13}. Regardless of the underlying 
cause, our motivation is to empirically choose galaxies at lower redshifts which are representative 
analogs of the higher-redshift population.  
In what follows, we make the assumption that the same evolution applies to both \ha\ and \oiii\ 
but different asumptions have also been implemented in this release of the MEx diagnostic code, 
as described in Appendix~\ref{app:altern}. 

It is possible that there is further evolution than fading of emission line luminosities associated 
with the decreasing star formation rates in galaxies.  For instance, \hii\ regions conditions may 
have been more extreme in the past, if struck by harder ionization fields and/or because of geometric 
constraints.  Theoretical predictions have been recently developed by, e.g., \citet{kew13a}.  We 
reserve discussion of these potential physical changes until Section~\ref{sec:theo}, but we note that 
our approach empirically accounts for the observed evolution of the mass-metallicity relation 
(Section~\ref{sec:MZ}).

The intermediate and high redshift samples were described in Section~\ref{sec:hizsamples}, and 
the line flux limit for each survey is given in Table~\ref{tab:samples}. 
For display purposes, we illustrate the prior sample corresponding to the line luminosity 
threshold at the median redshifts, i.e., the luminosity corresponding to the line flux detection 
limit minus the difference $\Delta L^*$ between the median redshift and that of the prior sample. 
The respective luminosity thresholds are $\log(L)>39.9, 40.2, 40.9, 40.8, 41.1$ for the samples at $\langle z 
\rangle = 0.45, 0.7, 1.4, 1.5, 2$.  An additional stellar mass limit is applied based on the minimum 
mass reached in each survey: $\log(M_{\star}[{\rm M_{\sun}}])>8.2, 8.6, 9.5, 9.0$ and $9.0$ for 
the samples at $\langle z \rangle = 0.45, 0.7, 1.4, 1.5$ and $2$, respectively.  
Combining this mass limit to the luminosity threshold, the prior samples can be used to predict the locus 
of higher-redshift galaxies on the BPT and MEx diagrams.

\begin{figure*}
\epsscale{1.} \plotone{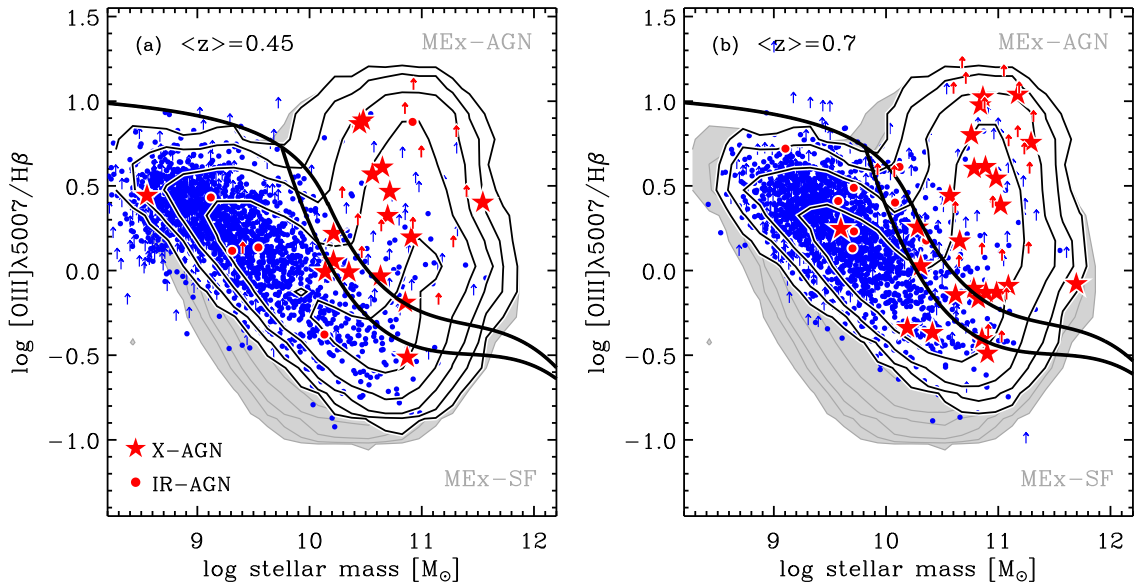}
\caption{
   MEx AGN diagnostic diagram applied to intermediate redshift galaxies:   
   (a) $\langle z \rangle=0.45$, 
   (b)  $\langle z \rangle=0.7$.
   The underlying gray shaded contours show the distribution of the full SDSS 
   prior sample, while the black contours include the priors selected to match the 
   detection luminosity threshold in each higher redshift subsample 
   (including mild $L^{\ast}$ evolution).  The observed high redshift galaxies 
   are shown with small filled circles, and highlighted with a large red circle (star)
   when there are known AGNs from mid-IR (X-ray). Lower limits on \oiii/\hb\ are indicated 
   with arrows, in red for mid-IR or X-ray AGNs, and in blue otherwise. 
   The agreement between the predictions from the prior samples and the observed galaxies 
   is generally good, especially with the X-ray AGN classification. 
   The demarcation curves were calculated for the detection limit 
   at the median redshift of each slice, and nicely separate the two branches seen in the 
   contours and in the individual points. Contours are logarithmically spaced (0.5~dex) with 
   the outermost contour corresponding to 10 galaxies per bin of 0.15\,dex$\times$0.15\,dex.
}\label{fig:hiz}
\end{figure*}

Figure~\ref{fig:hiz} illustrates the results for two intermediate redshift slices, at $\langle z 
\rangle = 0.45$ and $0.7$.  The bulk of the points are located 
within the contours predicted through the combined line detection limit and $L^*$ evolution.  AGNs identified 
independently through diagnostics at other wavelengths (X-ray and/or IR) are highlighted.  The agreement 
is especially good for AGNs thar are identified at least with X-ray signatures, with 81\% (80\%) 
lying in the MEx-intermediate or MEx-AGN regions at $z\sim0.45$ ($z\sim0.7$).  When including \oiii/\hb\ lower 
limits (red arrows), the agreement remains good with 86\% (87\%) at $z\sim0.45$ ($z\sim0.7$). However, there 
are more discrepancies with IR-only AGNs that are not also confirmed in X-rays (red circles).  Only 1/5 
at $z\sim0.45$ and 1/6 at $z\sim0.7$ lie in the MEx-intermediate or MEx-AGN regions.  The remainder  
is located in the MEx star-forming region, which indicate that they are mis-identified 
in either the MEx or the IR diagnostic (mid-IR colors).  
Mid-IR color diagnostics are known to suffer from contamination by star-forming galaxies in the AGN 
regions when deep IRAC observations are used \citep{bar06,don07,don12}, 
as is the case in this work.  Therefore, it is possible that some IR-only candidates 
are not truly AGNs, but their number is small enough that the overall AGN classification is 
satisfactory (76\% at $z\sim0.45$ and 79\% at $z\sim0.7$).  On the other hand, we do not expect to 
detect all MEx-AGNs with alternative methods as emission lines are the most sensitive probe 
and can reach much lower black hole accretion rates \citep{jun11,jun13}.

At $z>1$, we use three galaxy samples for which near-IR spectroscopy allows one to observe 
all the lines of the BPT-\nii\ (Table~\ref{tab:samples}). In each case, the same procedure is followed 
to empirically predict the loci of galaxies on both the BPT-\nii\ and the MEx diagrams.  The BPT 
-classification is also compared directly with that of the MEx diagram for the same galaxies. 
 
\begin{figure*}
\epsscale{1.} \plotone{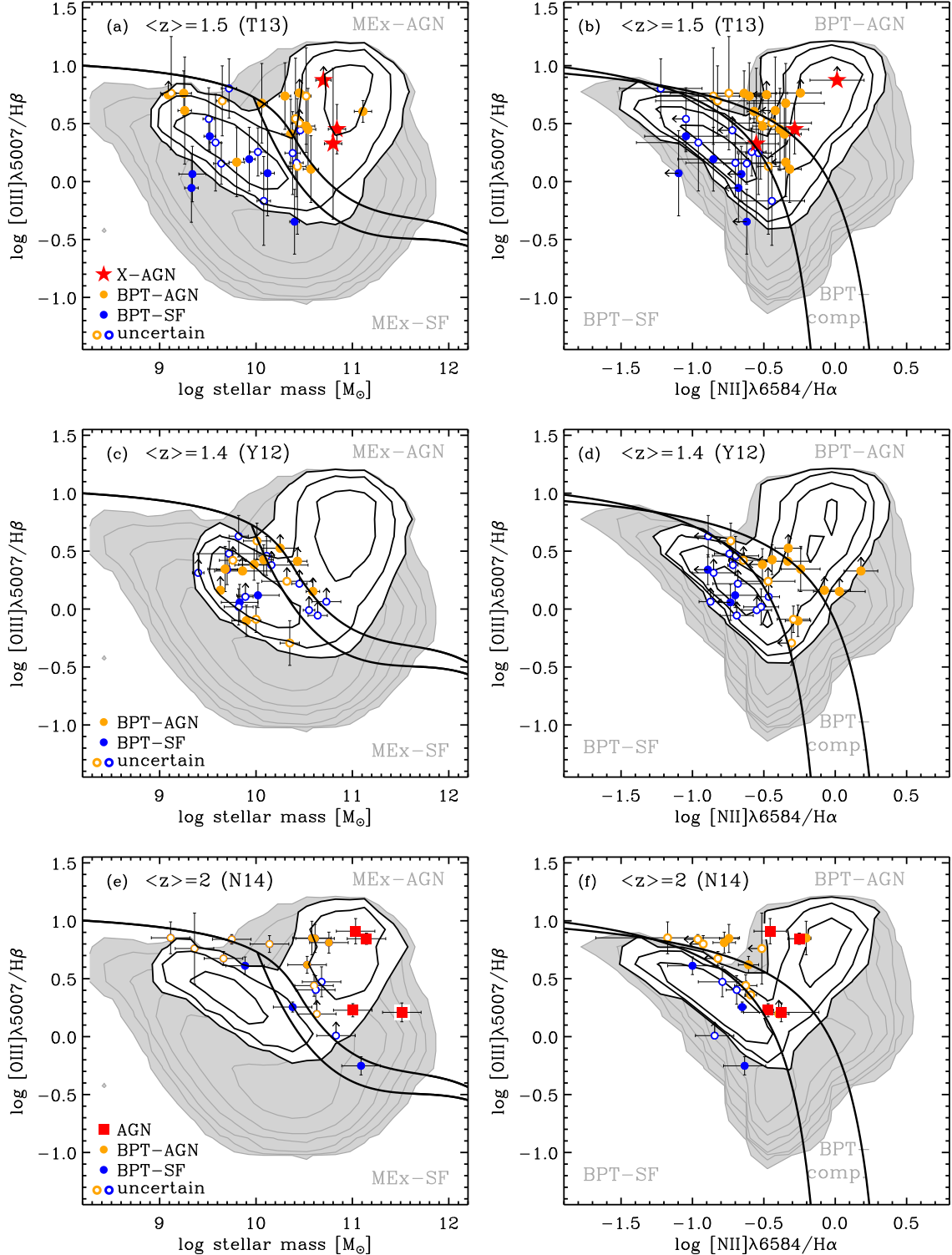}
\caption{
   AGN diagnostic diagrams applied to high redshift galaxies ($z\sim1.5-2$).  
   Three samples are shown (one per row).  In each row, the MEx is on the left-hand side 
   (panels a, c, e) while the BPT for the same galaxies is on the right-hand side 
   (panels b, d, f).  The top row illustrates the T13 sample at $z \sim 1.5$, 
   noting that X-ray identified AGNs are marked with red star symbols, and that points are 
   otherwise color-coded according to the BPT classes as labeled.  The uncertain classes 
   denote points which could be on either side of the Kauffmann line when accounting for 
   the error bars (open circles).  
   The second row shows the sample of Y12, with the same color coding except that X-ray 
   AGNs were discarded from their parent sample.  The third row shows the sample of N14 with 
   AGNs identified independently in red (from spatially-resolved line ratios or from  
   X-ray, UV, or mid-IR).  In all panels, 
   the underlying gray shaded contours show the distribution of $z\sim0$ emission-line 
   SDSS sample, while the black contours include the SDSS subsample selected to match the 
   detection luminosity threshold in each higher redshift subsample 
   (including the simple $L^*$ evolution). Contours are logarithmically spaced (0.5~dex) with 
   the outermost contour corresponding to 10 galaxies per bin of 0.15\,dex$\times$0.15\,dex.
}\label{fig:hizbpt}
\end{figure*}

The predictions from the SDSS priors appear suitable for the star forming 
branches of the BPT and MEx diagrams for both the $z\sim1,4$ and $z\sim1.5$ samples (Figure~\ref{fig:hizbpt}). 
The observed points lie mostly within the predicted contours, especially if we consider that the 
sizeable error bars broaden the true distribution. 
We remind that all three $z>1$ samples shown were not corrected for underlying Balmer absorption. 
Applying this correction would put their observed data points in slightly closer agreement with the predicted 
contours from SDSS galaxies by, e.g., shifting the points downward by $\sim$0.02 to 0.21~dex for the N14 sample 
(Section~\ref{sec:hizsamples}).
The T13 sample has larger uncertainties due in part to a full Monte Carlo simulation of the 
continuum fitting that yield $\sim30$\% larger error bars than the typical estimations, but also 
because low spectral resolution data were used for \oiii/\hb\ ratios. Despite the larger individual 
uncertainties, the ensemble of points satisfactorily constrain the bivariate distribution of 
selected sample.  Furthermore, when classifying individual galaxies on the BPT, we account for 
the uncertainties by marking cases which could be on either side of the \citet{kau03c} 
BPT dividing line as {\it uncertain}.

The two $z\sim1.5$ samples behave slightly differently with respect to one another on the AGN side. Some of 
the differences can be attributed to selection as Y12 rejected X-ray AGNs from their parent sample.  
A few of the BPT-AGNs from the T13 work lie between the SF and AGN branches while some of the BPT-AGNs 
from the Y12 study are located low on the right-hand branch, but the latter are mostly lower limits 
on the \oiii/\hb\ ratio, so they could still be fully consistent with the empirically predicted AGN branch. 

The predictions for the $z\sim2$ sample appear less consistent with the observations on the 
star-forming side of the diagrams. The $z\sim 2$ observations differ from both the $z\sim1,4$ 
and $z\sim1.5$ samples.  Perhaps $z>2$ marks a transition to strongly evolving ISM 
conditions in galaxies. However, it seems unlikely that there would be such a strong evolution 
during the 1~Gyr time span between $z=2$ and $z=1.5$, so the difference could instead be due to sample selection.  
One should keep in mind that the $z\sim2$ sample was selected for spatially resolved spectroscopy, 
with preference given to high-mass galaxies: 67\% (14/21) of galaxies have $M_{\star}>10^{10.5}$~\Msun\ 
in the N14 sample, while the corresponding fractions are 22\% (8/36; T13) and 15\% (4/27; Y12) for the 
$z\sim1.5$ samples. Therefore, the N14 sample is substantially different from the other two. 
At such high stellar masses ($>10^{10.5}$~\Msun), we predict that one needs to detect faint emission 
lines in order to probe star-forming galaxies.  As can be seen in Figure~\ref{fig:mex}, a luminosity 
threshold of $10^{40.2}$~\ergs still allows us to probe the star-forming branch on the MEx diagram, 
but this quickly vanishes as threshold line luminosities reaches $10^{40.3-40.4}$~\ergs.  Therefore,  
one needs more than three times fainter fluxes than achieved by typical surveys at $z\sim1.5-2$. 
In any case, given the various selection criteria and the small sample sizes, 
it is not surprising to see sample-to-sample variations.  However, we can still conduct internal comparisons 
such as comparing the BPT and MEx diagrams for a given galaxy sample.

To this aim, Figure~\ref{fig:hizbpt} is used to determine whether the BPT selection trends are analogous in the 
MEx diagram, and to compare the two AGN/SF classifications.  The K03 dividing line is used on the BPT-\nii\ 
in order to split BPT-AGN from BPT-SF galaxies.  
We incorporate information on the uncertainty of the classification for cases that overlap the dividing 
line when considering the measurement uncertainties ($1\sigma$ error bars or limits).  
Ignoring the cases with uncertain BPT classifications, the MEx tend to slightly under-predict the 
number of AGNs relative to the BPT, especially at low masses ($<10^{9.5-10}~M_{\sun}$).  On the other hand, 
there is virtually no contamination on the MEx-AGN side by BPT-SF galaxies in the $z\sim1.5$ samples, 
and a few possible BPT-SF in the MEx-intermediate zone ($0.3<P(AGN)<0.7$) in the $z\sim2$ sample.  
The number of $z\sim2$ star-forming galaxies in the N14 sample is too low to confirm whether this trend 
is significant. 

The fraction of galaxies with AGN according to the BPT and the MEx diagrams are listed 
in Table~\ref{tab:Agn}, where the MEx fraction is derived from the sum of AGN probabilities over the total 
number of galaxies.  On the BPT, we compute both the fraction of the most secure AGNs (solid yellow circles), 
which corresponds to a lower limit on the true AGN fraction, and the fraction assuming that all points above the 
Kauffmann line have an AGN (open and filled yellow circles).  In all cases, the 68\% confidence interval 
on the fractions is given based on Bayesian binomial statistics using algorithms from \citet{cam11}.  
The global AGN fractions from the MEx and BPT diagrams agree within the confidence intervals.

\renewcommand{\thefootnote}{\alph{footnote}}

\begin{deluxetable}{lccc}
\tabletypesize{\scriptsize}
\tablecolumns{4}
\tablewidth{0pc}
\tablecaption{Fraction of galaxies with AGN\label{tab:Agn}}
\tablehead{
   \colhead{Sample}  &  \colhead{MEx}  & \colhead{BPT}  & \colhead{BPT}  \\
   \colhead{ }  &  \colhead{ }  & \colhead{lower limit} & \colhead{all AGN} 
}
\startdata
$\langle z\rangle = 0.45$   &   $0.29\pm0.01$    &   \dots    &   \dots   \\
$\langle z\rangle = 0.7$    &   $0.24\pm0.01$    &   \dots    &   \dots   \\
$\langle z\rangle = 1.4$    &   $0.31^{+0.10}_{-0.07}$   &  $0.32^{+0.10}_{-0.07}$   &  $0.54^{+0.09}_{-0.09}$  \\
$\langle z\rangle = 1.5$    &   $0.42^{+0.08}_{-0.07}$   &  $0.39^{+0.09}_{-0.07}$   &  $0.58^{+0.08}_{-0.08}$  \\
$\langle z\rangle = 2$      &   $0.74^{+0.07}_{-0.11}$   &  $0.38^{+0.11}_{-0.09}$   &  $0.76^{+0.07}_{-0.11}$
\enddata
\tablenotetext{}{Columns: (1) Sample name; (2) AGN fraction from the MEx diagram using AGN probabilities; (3) AGN fraction from the BPT for only the most secure AGN (filled yellow circles on Figure~\ref{fig:hizbpt}); (4) AGN fraction from the BPT using the Kauffmann dividing line.}
\end{deluxetable}

\renewcommand{\thefootnote}{\arabic{footnote}}

We furthermore indicate AGNs that are identified independently from the global emission line 
ratios (red symbols in Figure~\ref{fig:hizbpt}). 
All seven of these objects are securely classified as AGN on the MEx diagram, and lie above the \citet{kau03c} 
line on the BPT (three of them are formally in the BPT-composite region and the remaining four in 
the AGN region above the \citet{kew01} line).  Relative to the other BPT-AGNs, the independently 
classified AGNs (in red) tend to reside in more massive hosts.  This was already observed at least 
for X-ray identified AGNs \citep[e.g.,][]{mul12,jun13}. The bias toward high stellar mass hosts 
for X-ray AGNs may be due to a selection effect associated with the more limited sensitivity 
of X-ray observations and the higher likelihood to detect AGNs with lower Eddington ratios in galaxies
hosting a more massive black holes, which themselves tend to be massive \citep{air12}.

The agreement between the BPT and the MEx diagrams is imperfect, with some BPT-AGNs 
lying on the MEx-SF side.  This occurs for 3 (or 4) among 18 BPT-AGNs for the T13 sample, 
and 4 (or 6) among 9 BPT-AGNs for the Y12 not counting uncertain classes and either excluding 
(or including) the lower limits on \oiii/\hb.  This could be due to the greater sensitivity of 
the BPT, reaching much lower accretion rates, therefore probing intrinsically weaker systems 
relative to X-ray or IR observations. Indeed, as we pointed out, none of the X-ray 
identified AGNs are missed by the MEx diagnostic diagram.  The presence of BPT-AGN in the 
MEx star-forming region may also reflect AGN incompleteness for hosts with low stellar masses 
($<10^{9.5-10}$~\Msun), or potential mis-classification on the BPT diagram if it evolved 
with redshift \citep{kew13a,kew13b}.

\subsection{Comparison with Theoretical Evolution Models}\label{sec:theo}

Recently, \citet{kew13a} presented a theoretical approach to 
predicting the location of higher-redshift galaxies on the BPT diagnostic diagram. 
Their framework is based on a description of two sequences on 
the BPT-\nii\ diagram: an {\it abundance sequence} corresponding to the star-forming 
branch, and a {\it mixing sequence} where the AGN contribution to the emission lines 
rises toward the upper right part of the diagram. \citet{kew13a} use the photoionization 
code MAPPINGS IV \citep{dop13} with input Starburst99 stellar population models radiating 
on ISM with varying metallicities to define the abundance sequence. The AGN contribution 
is added with emission lines from dusty Narrow-line region (NLR) models calculated 
with MAPPINGS III as described by \citet{gro04a}. An increasing AGN contribution forms 
the mixing sequence\footnote{\citet{kew13a} also explored slow shock models and found them 
to predict line ratios that are distinct from those produced by AGN photoionization 
(their Section~5.1).}.

\begin{figure*}
\epsscale{1.} \plotone{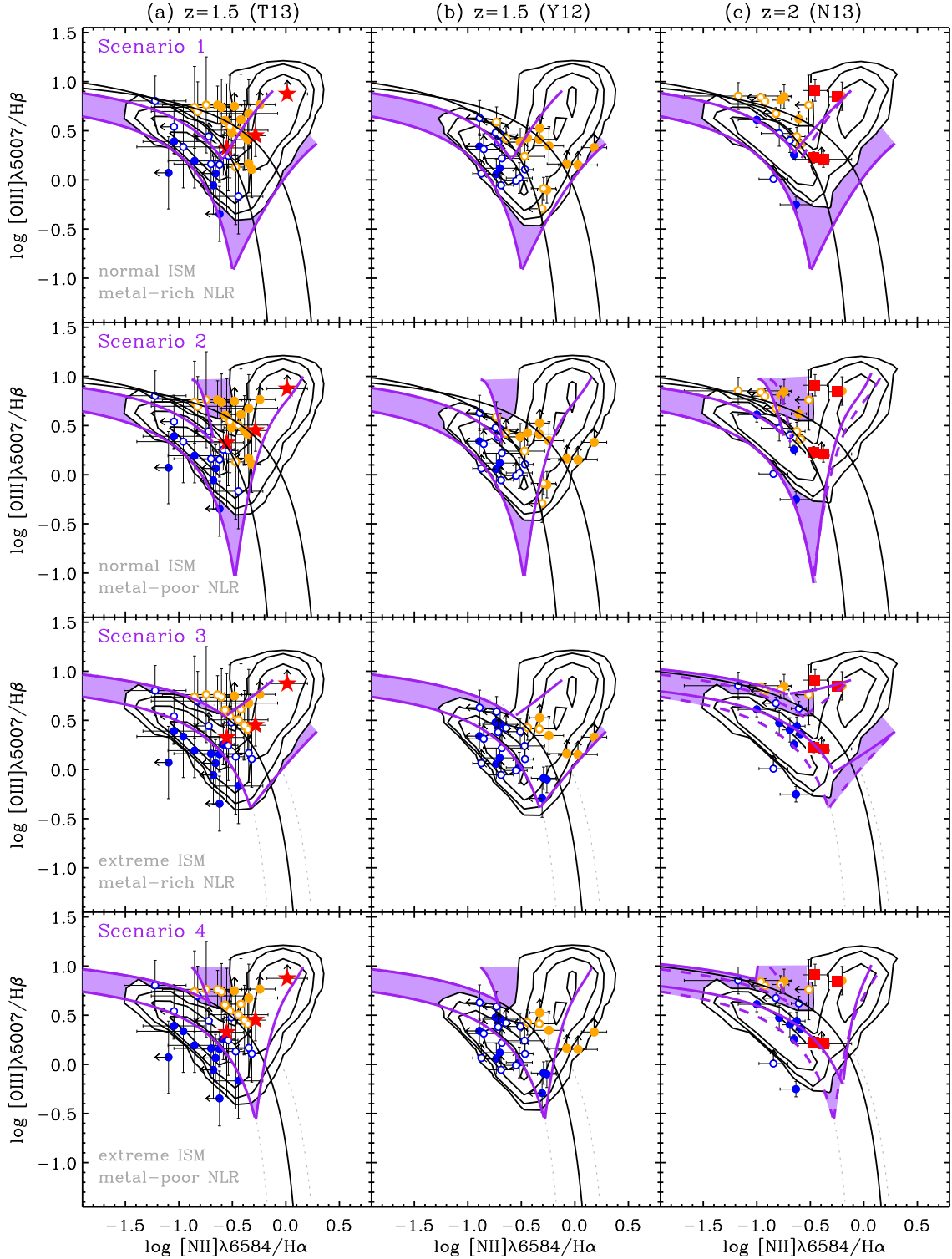}
\caption{
   BPT-\nii\ diagrams for the three $z>1$ samples, in columns (a), (b), and (c), as labeled.
   The SDSS-based empirical predictions (black contours) are compared with the theoretical 
   predictions of the scenarios from \citet{kew13a} (purple regions).  Each row corresponds 
   to a scenario from 1 (top) to 4 (bottom), each with different assumptions for the evolution 
   in ISM conditions and NLR metallicity.  
   For the $z\sim2$ sample, the evolutionary trends are shown at $z=1.5$ (dashed lines) 
   and $z=2.5$ (solid lines), bracketing the redshift range of the observations.  
   Plotting symbols are identical to Figure~\ref{fig:hizbpt}. Contours are logarithmically spaced (0.5~dex) with 
   the outermost contour corresponding to 10 galaxies per bin of 0.15\,dex$\times$0.15\,dex.
}\label{fig:allsce}
\end{figure*}

Kewley and collaborators explored four different evolutionary scenarios:
\begin{enumerate}
\item[1.] Normal ISM and metal-rich AGN NLR at high $z$
\item[2.] Normal ISM and metal-poor AGN NLR at high $z$
\item[3.] Extreme ISM and metal-rich AGN NLR at high $z$
\item[4.] Extreme ISM and metal-poor AGN NLR at high $z$  
\end{enumerate}

In the first two cases, the abundance sequence remains unchanged with redshift.  
Purely star-forming galaxies are predicted to move along the sequence as their 
gas-phase metallicity evolves. According to the last two scenarios, star-forming 
galaxies have more extreme ISM conditions, with a higher ionization parameter, 
higher electron densities, higher star formation surface densities or a combination.  
In these cases, the locus of the abundance sequence shifts toward higher \oiii/\hb\ 
at higher redshifts. 

For each set of ISM conditions, there are two alternative scenarios for the abundance 
of the AGN NLRs.  The NLR gas is assumed to either be enriched early, meaning that 
the NLRs are metal-rich across the full range of redshifts considered, or to have a 
gas-phase abundance that changes with cosmic time following the evolution of the 
$M_{\star}-Z$ relation for star-forming galaxies. A lower NLR metallicity yields to lower \nii/\ha\ 
ratios \citep{gro06}, and therefore changes the mixing sequence.  These four scenarios 
are compared to observations of galaxies at intermediate to high redshifts by \citet{kew13b}.  

Figure~\ref{fig:allsce} illustrates a comparison between our empirical approach and 
the four scenarios developed by K13, applied at $z>1$.  At first glance, the $z\sim0$ empirical 
predictions (black contours) appear intermediate between scenarios 1 and 2, 
with normal ISM conditions.  
However, an important distinction is that the empirical method predicts the absence 
of galaxies with comparatively low \oiii/\hb\ for the high redshift samples, due to 
line luminosity detection threshold.  
In their initial comparison with observations, \citet{kew13b} have not included selection 
effects caused by the non-detection of intrinsically faint lines at higher redshifts.  
Including those effects would namely trim the region predicted to reach low 
\oiii/\hb\ values according to Scenarios~1 and 2, thus increasing the lower envelopes 
of the K13 models with normal ISM conditions.  

If the empirical predictions from the selected $z\sim0$ SDSS subsamples favor Scenarios~1 
and 2, what is the preferred scenario according to the observed high-redshift galaxies?
Because of the large sample-to-sample variation between the observations, it is not 
obvious what scenario is prefered (Figure~\ref{fig:allsce}).  
On the star-forming side, both the $z\sim1.4$ and $z\sim1.5$ samples 
are fully consistent with the empirical predictions from the matched $z\sim0$ galaxies.   
Therefore, the locus of those $z\sim1.5$ galaxies is better represented by normal ISM 
conditions (with selection effects) rather than the strong ISM evolution of some of the 
K13 models (scenarios 3 and 4). The case of the N14 sample is less clear because of the 
dearth of low and moderate mass galaxies (Figure~\ref{fig:hizbpt}), and therefore the predicted 
small number of star-forming galaxies in that sample (Section~\ref{sec:hiz}). 
It remains possible that this sample favors a more strongly evolving ISM at 
$z>2$, which could be supported by even higher redshift results at $z\sim3$ 
\citep[e.g.,][]{hol14}.

Regarding the AGN branch, the T13 and N14 samples (columns a and c of Figure~\ref{fig:allsce}) 
appear to be better represented with metal-poor NLR on the AGN side (Scenarios 2 and 4).   
The Y12 sample lies between the predictions from metal-poor and metal-rich NLRs, but also 
includes lower limits on \oiii/\hb, which prevent us from knowing the true distribution of 
the small number of AGNs. We also recall that Y12 excluded AGNs that were a priori identified 
from their parent sample, and so this sample provide us with a less stringent contraint on the AGN side. 
Overall, the $z\sim1.5$ data may favor Scenario~2, with normal ISM in galaxies and 
metal-poor NLRs at higher redshifts, although we note that larger samples are needed to 
confirm this conclusion, and whether there is a transition to stronger ISM evolution 
at $z>2$.

\section{Discussion}\label{sec:discu}

\subsection{Mass-Metallicity Relation}\label{sec:MZ}

Several studies have reported evolution of the stellar mass-metallicity ($MZ$) relation for galaxies \citep{sav05,erb06,zah13}, 
or along a plane in the $M_{*}-SFR-Z$ space \citep{man10,lar10,yat12}.  
There are variations in the details of this evolution and how it may vary with galaxy stellar masses, but 
the general sense is that at a given stellar mass, lower redshift galaxies have higher gas-phase 
metallicities than their higher redshift counterparts.  This could indicate the global enrichment  
in galaxies as they evolve and form stars.  

\begin{figure*}[bht]
\epsscale{1.0} \plotone{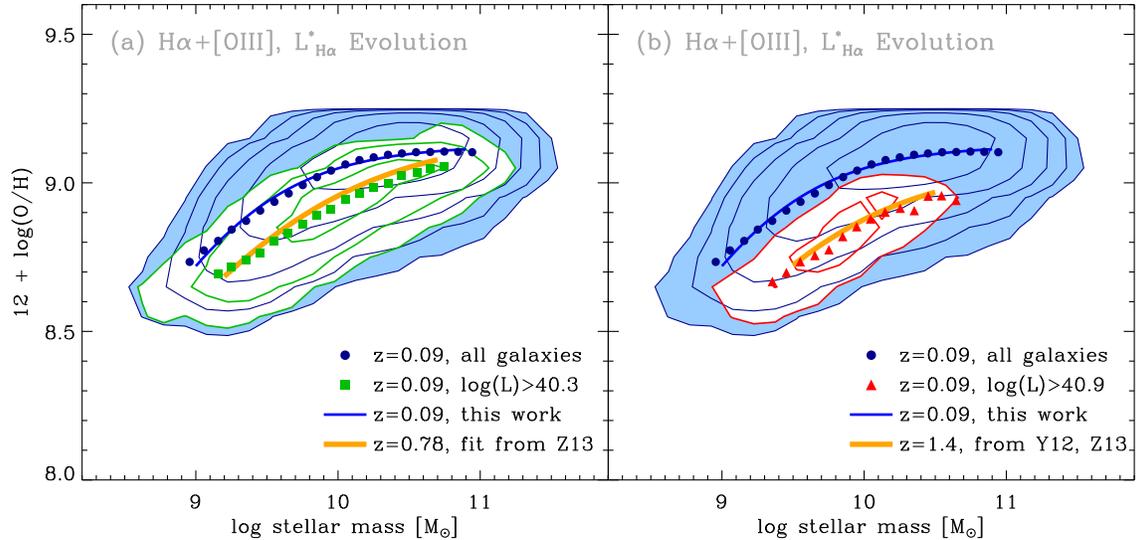}
\caption{
   $MZ$ relation from SDSS observations at low-redshift (this work) and from observations compiled by \citet{zah13}, 
   who fitted DEEP2 $z\sim0.8$ and Y12 $z\sim1.4$ results (thick orange curves).
   In all panels, the blue logarithmically-spaced contours show the full SDSS prior galaxy sample, 
   with filled circles marking the median metallicity in stellar mass bins of 0.1~dex.  The blue line shows the 
   fitted relation using the functional form introduced by \citet{mou11}.
   Green and red contours correspond to a SDSS subsample with \ha\ and \oiii\ lines above the labeled luminosity. 
   The median metallicity in stellar mass bins of 0.1~dex is shown with filled symbols.  The choice of the threshold 
   luminosity is made according to the fiducial scenarios, where the luminosity detectability threshold 
   is lowered by the same amount as $L^*$(\ha) fades between the redshifts of interest:  (a) $z\sim0.8$ and (b) $z\sim1.4$, 
   and the prior sample at $z\sim0.09$. In this case, the predicted contours and median values agree remarkably well with 
   the observed relations compiled by Z13.  Alternative scenarios and more details are included in 
   Appendix~\ref{app:altern}. Contours are logarithmically spaced (0.5~dex) with 
   the outermost contour corresponding to 50 galaxies per bin of 0.15\,dex$\times$0.15\,dex.
}\label{fig:MZ}
\end{figure*}

The revised MEx demarcations and probabilities now include metallicity evolution indirectly through the use of 
line luminosities to build a prior sample.  As we show below, together with the stellar mass, emission line 
luminosities can trace metallicity to some extent. This is similar, though not identical, to previous work 
defining the Fundamental Metallicity Relation \citep{man10,lar10,yat12,cre12}. 
In the latter, the SFR is used as a third parameter to define 
a plane while here we use both \ha\ luminosity, a tracer of the SFR \citep{ken98}, and \oiii, which depends 
more directly on the gas-phase metallicity. We also {\it fade} the line luminosity threshold by the corresponding 
fading of the knee of the \ha\ luminosity function. 

We test directly our selection method against observed $MZ$ relations compiled by \citet[hereafter Z13]{zah13}. 
These authors fitted the functional form defined by \citet{mou11} to datasets in five redshift slices, 
including an intermediate-redshift DEEP2 sample ($z\sim0.8$), 
and the Y12 sample ($z\sim1.4$), which overlap with the present study. 
These $MZ$ relations are compared to our empirical predictions from 
SDSS prior samples. We compute metallicities following the same method as Z13 to facilitate 
a direct comparison.  As in that study, we apply the KK04 calibration \citet{kob04} to the required 
\oii, \hb, and \oiii\ emission lines. Line fluxes were first corrected for dust attenuation by 
measuring the Balmer decrement and applying the \citet{cal00} dust attenuation curve with $R_V=4.05$, 
assuming an intrinsic ratio of \ha/\hb=2.86 \citep{ost06}.

In Figure~\ref{fig:MZ}, we show predicted contours using our 
fiducial approach (\ha\ and \oiii\ selection above the luminosity threshold that includes $L^*_{\ha}$ evolution). 
There is a good global agreement between predicted contours and analytical fits to observations reported 
by Z13, indicating that the observed $MZ$ evolution is a built-in feature of our method. Empirical predictions 
were also made for two alternative scenarios (Appendix~\ref{app:altern}), namely to compare with samples 
selected from a single line: \oiii\ or \ha, but also to consider null evolution scenarios where the 
line luminosity detection limits are used as is.

We note that the empirical prediction of the Y12 sample (red contours on panel b) show a cutoff at 
high masses ($>10^{10.5}$~\Msun).  While galaxies with these high masses and low metallicities may 
be more common at higher redshifts.  This is likely a true evolutionary trend, and could 
indicate that most massive galaxies have had time to enrich their ISM by $z<0.2$ and that the galaxies 
that remain with low metallicities at this more recent epoch have lower stellar masses.

\subsection{Is there evolution in the emission-line diagnostics?}\label{sec:shift}

In Section~\ref{sec:detlim}, we investigated the consequences of increasing emission-line 
luminosity limits to select subsamples from $z\sim0$ SDSS emission-line galaxies, and the 
resulting bivariate distributions on the BPT and MEx diagrams. We found that the division 
line developed by \citet{kau03c} still appears to be applicable to divide the two branches 
defined by the data at all the luminosity thresholds tested, albeit a small shift of $<$0.2dex 
toward higher \nii/\ha\ and/or \oiii/\hb\ could be possible at the highest line luminosities.
This suggests that the luminosity dependence would more strongly affect the locus of the sample 
than the definition of the dividing lines. 
In Section~\ref{sec:hiz}, we compared $z\sim1.5-2.5$ observations with empirical predictions 
from SDSS subsamples selected to have equivalent line luminosity limits.  We found a general 
but imperfect agreement between the observations and matched SDSS prior samples 
on the BPT diagrams.  Namely, some galaxies from the T13 and N14 samples 
are located between the two empirically predicted BPT branches on the right-hand side panels of 
Figure~\ref{fig:hizbpt}.  This feature could be due to lower NLR metallicity, such as 
predicted by K13 with their Scenario~2 (second row of Figure~\ref{fig:allsce}). 
Alternatively, it could be due to AGN host having higher SFRs at higher redshifts, which 
would boost \ha\ more strongly than \nii\ and dilute the AGN signatures.

Lower NLR metallicites and/or higher SFRs at higher redshift imply a potentially greater 
fraction of AGNs in the BPT-composite region (between the \citealt{kau03c} and \citealt{kew01} 
lines) and even toward the top of the star-forming branch.  This trend could explain the presence 
of X-ray (or otherwise securely-identified) AGNs on the boundary between the composite and 
star-forming regions of the BPT-\nii\ diagram in the T13 and N14 samples (red symbols in 
Figure~\ref{fig:hizbpt}) and the presence of data points to the left of the AGN branch on the 
BPT diagram. As a consequence, AGN samples should include BPT-composites 
to improve their completeness \citep[also see][]{tro11}. Conversely, purely star-forming 
galaxy samples may be harder to obtain given the higher risk of including AGN contaminants 
when NLR metallicities are lower. A potential solution may be to combine both the BPT and MEx 
classification schemes in such ambiguous cases.

How do these trends compare on the MEx diagram?  In contrast to the BPT diagram, the 
splitting of the two branches traced by $z\sim0$ SDSS samples on the MEx diagram 
shows an obvious offset with increasing line luminosity limits (Figure~\ref{fig:mex}).  
This offset corresponds to changes in the AGN probabilities as a function of line luminosity 
limits, with the transition region shifting toward higher stellar masses with 
increasing line luminosity limits (fitted in Appendix~\ref{sec:offset}). The physical interpretation 
depends on the nature of the main ionizing source.  For purely star-forming galaxies, brighter 
lines correspond to lower metallicity and/or higher SFRs, and therefore higher \oiii/\hb\ ratios.  

However, on the AGN side, brighter lines imply higher accretion rates 
onto black holes (traced by \oiii) as well as higher SFR of the hosts (traced by \ha).  
Both of these criteria will favor high-mass hosts. First, massive galaxies are more likely to 
host high-mass black holes \citep[e.g.][]{mag98,geb00,fer00}, which will be brighter than 
lower mass black holes for a given Eddington ratio distribution \citep[also see][]{air12}.  
Second, high mass galaxies are more likely to have high SFRs, according to the 
$M_{*}-$SFR sequence for star-forming 
galaxies\footnote{The emission line selection will favor star-forming or active galaxies 
over truly passive systems which do not follow the $M_{*}-$SFR sequence.  This bias will be 
increasingly important as the emission line luminosity limit increases, therefore increasing 
the likelihood that the selected galaxies are star-forming and not passive.} 
\citep{bri04,noe07,elb11}.
Therefore, these trends likely combine to shift the star-forming galaxies/AGN division toward 
higher stellar masses.  Regardless of the underlying interpretation, the empirical offset 
has been calibrated, allowing one to trace the MEx demarcation lines for various line sensitivity 
limits.  The MEx AGN probabilities can be calculated taking into account both the 
survey detection limit, and the individual redshift of each galaxy (which will determine the 
amount of $L^*$ evolution).

While the $z>1$ samples on the BPT diagram suggested offsets of AGN NLRs toward lower 
metallicities (i.e., lower \nii/\ha\ ratios), the MEx diagram is largely insensitive to such 
an effect, as the \oiii/\hb\ would vary in the opposite direction, \emph{helping} the AGN 
selection.  Instead, the MEx diagnostic has the caveat of being incomplete to select AGNs 
at low stellar masses ($<10^{9.5-10}$~$M_{\sun}$).  This limitation depends slightly 
on line luminosities with this new approach (galaxies with more luminous emission liness 
can only be recognized as AGNs in higher mass hosts relative to galaxies with fainter lines).  
This limitation may not be very severe if AGNs in low-mass hosts are rare \citep{bel11,tan12}, 
and/or if their relative importance was lower at higher redshift, as would be the case if AGN 
activity followed the downsizing phenomenon \citep[e.g.,][]{bar05,kel13,hir13}.  The situation is different if 
one is particularly interested in low-mass AGN hosts, low-mass black holes and black hole 
seeds \citep{gre07,bar08,dia13,rei13}.

\subsection{Lower NLR metallicity at higher redshifts: gas-poor hosts or metal dilution?}\label{sec:interp}

NLRs tend to be metal-rich in nearby galaxies, and while there is evidence for low-metallicity 
AGNs, those systems are rare \citep{gro06}.  In this study, we find tentative evidence that higher redshift 
AGNs may be less chemically enriched than their local counterparts.  We discuss a few physical mechanisms that could 
explain this trend, in light of our recent understanding of high-redshift star-forming galaxies.

As also mentioned by \citet{kew13a}, emission lines from NLRs may trace gas closer to galaxy nuclei 
than global galaxy-scale spectra. Negative metallicity gradients could therefore play a role in explaining 
that NLRs are typically more metal-rich than the surrounding galaxy ISM at larger scales. In this view, 
if disk galaxies start with flatter gradients at higher redshifts, the NRLs in these hosts would also 
exhibit more metal-poor characteristics. Conversely, if the nuclei of galaxies enrich on very short timescales, 
then the NLRs would be metal-rich already in higher redshift systems, meaning that their hosts had steeper 
metallicity gradients.  Thus, NLR metallicities could trace whether host galaxies have had time to enrich 
at least their central regions, and whether disks grow inside-out \citep[e.g.,][]{jon13}. However, this simple 
picture may not hold during galaxy mergers because metal-rich nuclei could be diluted by inflows 
of more pristine gas brought in from the outskirts or from satellite galaxies.  Galaxy mergers 
and interactions have indeed been reported to exhibit flattened or inverted metallicity gradients 
\citep[e.g.][]{kew10,rup10,que12}, though major merger events only account for a small fraction of 
the total star-forming and AGN galaxy population at a given time.

Here, we discuss one more possibility related to violent disk instabilities \citep[VDIs; e.g.,][]{bou07,dek09}.  
High-redshift disk galaxies have been observed to have high gas fractions at $z>1.5$ \citep{dad10,tac10}, 
and clumpy appearances \citep[e.g.][]{elm07,elm09} that distinguish them from typical star-forming 
disk galaxies observed at low redshifts, and that are interpreted as observational signatures of 
VDIs.  These instabilities are predicted to be ubiquitous at $z>1-2$, and to generate 
inflows toward the central regions \citep{kru10,bou11}. 
These inflows can bring metal-poor gas in the viscinity of active BHs and result in lower metallicity NLRs.  
While VDI clumps are themselves star-forming and produce metals, they undergo outflows \citep{gen11}
as they migrate inwards but also accrete gas from their diffuse surrounding \citep{dek13,bou14}.  This 
could potentially maintain somewhat lower metallicities for these clumps, or the general turbulence 
could contribute to erase or flatten metallicity gradients \citep[e.g.,][for an example case]{que12}. 
The details and timescales are still uncertain, but we speculate that VDIs could play a role 
in determining the observed metallicities of NLR gas at higher redshifts ($z>1$).

\subsection{Comparison with previous MEx diagram results}\label{sec:comp}

In this work, we have revised the MEx demarcation and probability calculations to use 
the SDSS DR7 sample (Section~\ref{sec:revised}).  We have then implemented changing MEx 
demarcation and probabilities as a function of the effective luminosity threshold for emission 
line detection (applied to \ha\ and \oiii).  How do these revisions compare to other studies 
of the MEx at high-redshift from the literature?

On the one hand, T13 had found a good agreement between the BPT classification and the original MEx 
dividing curves from J11, and concluded that the original MEx classification were valid for their 
sample at $z\sim1.5$.  The revised demarcations are now shifted slightly upwards at low 
masses based on the low-redshift calibration with SDSS DR7 (Figure~\ref{fig:newmex}). 
This feature increases the number of low-mass BPT-AGNs that lie in the MEx-SF region at the 
top left of the star-forming branch and slightly worsens the agreement.  

On the other hand, N14 noted that the original MEx demarcations should be displaced to higher 
stellar masses in order to improve the agreement between the BPT and MEx diagrams at $z\sim2$.  
In this case, our revised MEx dividing curves improve the agreement and mitigate the need for such a large 
shift. Instead, the revised MEx demarcation lines introduced here appear to be applicable 
out to $z\sim2$ given the current observational constraints, and given the limitations of the 
uncertain BPT classes for galaxies with measurement errors spanning the AGN/SF demarcation, 
which were not taken into account by N14.  More recently, \citet{hen13} also suggested that the 
MEx demarcation should shift to higher stellar masses at higher redshifts, and suggested a 
1~dex shift for their galaxy sample at $1.3<z<2.3$.

The empirical approach presented in this Paper lies between these two conclusions, i.e., it includes 
a line luminosity dependence that mimic high-redshift galaxies on line ratio diagnostics, and accounts 
for $MZ$ relation evolution (Section~\ref{sec:MZ}) but the mass offsets on the MEx doagnostic diagram 
are generally not as extreme as those suggested by N14 and \citet{hen13}.  
A direct consequence of using the revised MEx diagnostic at higher redshifts will be to yield slightly 
lower AGN fractions compared to the original J11 version.  This was noted by \citet{mig13} in their 
work comparing AGNs identified from \nevlam\ and X-rays in zCOSMOS, with the MEx diagnostic among others.

The MEx demarcation lines can be further tested in future investigations by using the publically available MEx 
probability calculation code\footnote{https://sites.google.com/site/agndiagnostics/home/mex}, 
with optional emission line detection limits tailored to each survey, and with or without prescriptions 
for $L^*$ evolution to the redshift of each individual galaxy. It will also be informative to push the 
analysis to yet higher redshift, with new samples becoming available \citep[e.g.][]{hol14}, which will be 
a follow-up work to this article.

In Appendix~\ref{app:stac}, we present 
a example application to intermediate-redshift ($0.2<z<0.8$) galaxy samples from the stacked spectral 
analysis of \citet{vit13}. We have also implemented alternative selections based on a single emission line, 
\ha\ or \oiii, in the public distribution of the code (Appendix~\ref{app:altern}).

\section{Summary}\label{summ}

Galaxies at higher redshifts appear offset from the locus of low-redshift galaxies on emission-line 
diagnostic diagrams such as the BPT and MEx diagrams.  These two planes share a common vertical axis 
and behave similarly but with noticeable differences.  In this Paper, we have investigated the cause of this 
apparent redshift-dependent offset, improved the applicability of the MEx diagram to a range of redshifts, 
and demonstrated the crucial importance of taking into account selection effects due to emission line detection limits.

Our main results are:

\begin{enumerate}

\item[1.] We have revised the $z\sim0$ demarcations of the MEx diagnostic diagram with a $0.04<z<0.2$
  prior sample of emission-line galaxies selected from SDSS DR7, a superset of the SDSS DR4 sample that 
  was used in the original definition of the MEx by J11.

\item[2.] Imposing a minimum line luminosity to \ha\ and \oiiilam\ affects 
  the bivariate distribution of the galaxies on the BPT and MEx diagrams.  With increasing line luminosity, 
  the shift is in the sense of higher \oiii/\hb\ ratios in both cases (Figures~\ref{fig:bpt} and \ref{fig:mex}), 
  and toward lower \nii/\ha\ for the AGN branch of the BPT.  Therefore, 
  we find that selection effects applied to $z\sim0$ samples mimics the high values of \oiii/\hb, and 
  (among AGN) comparatively lower \nii/\ha\ ratios seen in galaxy samples at $z>1$. 

\item[3.]  In the case of the MEx diagram, the splitting between the star-forming branch and AGN branch 
  -- where the SF/AGN classification comes from the BPT diagram -- occurs at increasingly higher stellar mass 
  as the cutoff line luminosity is raised. We thus develop a line detection-limit dependent MEx diagnostic diagram. 

\item[4.] At $z<1$, optical spectra of $\sim$3400 galaxies with \oiii\ and \hb\ are used on the MEx diagram.  
  The line luminosity threshold was calculated in two redshift slices, $0.3<z<0.6$ and $0.6<z<1$, using the formal 
  flux detection limit of the surveys, corrected for a mild luminosity evolution of the population 
  ($L^*_{\ha}$ evolution). These luminosity limits of $10^{40}$ and $10^{40.4}$~\ergs empirically predict the locus 
  of these respective sample on the MEx diagram (Figure~\ref{fig:hiz}).

\item[5.] At $z\sim1.5-2$, we gather 
  three published galaxy samples with near-IR spectroscopy for which all four BPT lines are covered, enabling us to 
  apply both the BPT and MEx diagnostics (Figure~\ref{fig:hizbpt}). The prior samples predict well the 
  allowed region on the diagrams, without invoking evolving \hii\ conditions out to $z\sim1.5$. 
  The case at $z\sim2$ is less clear due to small sample size, a high fraction of massive ($>10^{10.5}$~\Msun) galaxies, 
  and ambiguous classification on the BPT diagram (conflicting classes when accounting for line ratio uncertainties).

\item[6.] A comparison with theoretical predictions by \citet{kew13a} and observations at $z\sim1.5$ suggests 
  that the favored scenario for star-forming galaxies may be {\it normal} \hii\ region conditions when 
  one also accounts for selection effects.  The latter tend to prevent the detection of galaxies with 
  low \oiii/\hb\ ratios. There may be transition at $z>2$ toward more strongly evolving 
  ISM conditions but future work is needed to confirm this possibility.

\item[7.] On the AGN side, the data appear to favor K13 scenarios with varying metallicity of 
  narrow-line regions photoionized by AGN, with lower metallicities at higher redshifts. 
  The observed sample-to-sample variation at $z>1$, and the sizeable error bars, do not allow us to firmly 
  rule out the metal-rich NLR scenario. 

\item[8.] Selecting emission lines based on detectability at higher redshifts 
  allow us to reproduce remarkably well the observed evolution of the $MZ$ relation
  (Figure~\ref{fig:MZ}).  Thanks to this feature, the locii of prior SDSS samples 
  on the MEx diagram and associated AGN probabilties include this reported $MZ$ evolution.  
  We find that the slope and/or offset of observed $MZ$ relations may 
  depend on sample selection and emission-line detection limits (Figure~\ref{fig:MZOiii}).

\item[9.] In addition to a fiducial method where we require both \oiii\ and \ha\ lines 
  to be above a common threshold luminosity, we also implement selection and evolution scenarios based on 
  a single emission line and/or based on different evolutionary schemes (including a null evolution; 
  Appendix~\ref{app:altern}).  These scenarios can offer a better description of observed 
  high-redshift samples than using the full SDSS sample, and generally stress the importance of 
  accounting for sensitivity limits in emission-line studies.

\end{enumerate}

Larger galaxy samples with a well-understood selection function and deep spectral coverage are required 
to better disentangle between ISM evolutionary scenarios.  However, when such samples are available, 
care should be taken to include selection effects in order to avoid ruling out a scenario based on 
the absence of observed galaxies in a region that is beyond the detectable limit, such as low \oiii/\hb\ 
ratios on AGN diagnostic diagrams, and correspondingly a high gas-phase metallicities for 
galaxy abundance surveys relying on \oiii\ detection. 
Properly accounting for selection bias will be crucial to unravel the 
underlying physics explaining the locus of galaxies on nebular line diagrams.
The empirical approach and publicly available code presented here should be applicable to future 
medium and large-scale NIR spectroscopic surveys, with facilities such as Magellan/MMIRS, 
Keck/MOSFIRE, Subaru/FMOS, VLT/KMOS, Gemini/Flamingos-2, and JWST/NIRSpec as well as Euclid.

\acknowledgments 

The authors thank the anonymous referee for comments that greatly improved this article, and that motivated 
the development of a much more versatile tool. 
The authors also thank K. Yabe for kindly sharing the stellar masses and line ratios for their galaxy sample \citep{yab12}, 
and M. Vitale for generously sharing their stacked spectra results \citep{vit13}.  
SJ would like to acknowledge useful discussions with L. Kewley, M. Mignoli, G. Cresci, M. Brusa, and M. Sarzi 
as well as earlier discussions with C. Scarlata and H. Teplitz.
SJ and FB acknowledge support from the EU through grant ERC-StG-257720.
SC acknowledges support from the European Research Council via an Advanced Grant under grant agreement no. 321323.

Funding for the SDSS and SDSS-II has been provided by the Alfred P. Sloan Foundation, the Participating Institutions, the National Science Foundation, the U.S. Department of Energy, the National Aeronautics and Space Administration, the Japanese Monbukagakusho, the Max Planck Society, and the Higher Education Funding Council for England. The SDSS Web Site is http://www.sdss.org/.

The SDSS is managed by the Astrophysical Research Consortium for the Participating Institutions. The Participating Institutions are the American Museum of Natural History, Astrophysical Institute Potsdam, University of Basel, University of Cambridge, Case Western Reserve University, University of Chicago, Drexel University, Fermilab, the Institute for Advanced Study, the Japan Participation Group, Johns Hopkins University, the Joint Institute for Nuclear Astrophysics, the Kavli Institute for Particle Astrophysics and Cosmology, the Korean Scientist Group, the Chinese Academy of Sciences (LAMOST), Los Alamos National Laboratory, the Max-Planck-Institute for Astronomy (MPIA), the Max-Planck-Institute for Astrophysics (MPA), New Mexico State University, Ohio State University, University of Pittsburgh, University of Portsmouth, Princeton University, the United States Naval Observatory, and the University of Washington.

This work is based in part on observations made with the {\it Spitzer} 
Space Telescope, which is operated by the Jet Propulsion Laboratory, 
California Institute of Technology under a contract with NASA,  
and on observations made with the {\it Chandra} X-ray Observatory, operated by the Smithsonian 
Astrophysical Observatory for and on behalf of NASA under contract NAS8-03060.
The TKRS survey was funded by a grant to WMKO by the National Science Foundation's Small Grant for Exploratory Research program. The resulting data are a combined set of data obtained by both Team Keck and their collaborators at the University of Hawaii/Institute for Astronomy.

This study makes use of data from AEGIS, a multiwavelength sky survey conducted with the Chandra, GALEX, Hubble, Keck, CFHT, MMT, Subaru, Palomar, Spitzer, VLA, and other telescopes and supported in part by the NSF, NASA, and the STFC.
Funding for the DEEP2 survey has been provided by NSF grants AST95-09298, AST-0071048, AST-0071198, AST-0507428, and AST-0507483 as well as NASA LTSA grant NNG04GC89G. The analysis pipeline used to reduce the DEIMOS data was developed at UC Berkeley with support from NSF grant AST-0071048. 

{\it Facilities:} \facility{Spitzer (IRAC)}, \facility{Keck (DEIMOS)}, \facility{HST (ACS)}, \facility{Chandra (ACIS)}

\appendix
\section{A. Revised Emission Line Flux Uncertainties for SDSS DR7}\label{app:err}

The MPA/JHU analysis includes uncertainty scaling factors obtained from SDSS DR4 by 
comparing the spread between individual measurements for duplicate observations of the same 
objects with the formal errors quoted in their emission line catalogs \footnote{The DR4 
uncertainty scaling factors and the method used to compute them are available at the following URL: 
http://www.mpa-garching.mpg.de/SDSS/DR4/raw\_data.html}.
However, this exercise was not repeated with the DR7 spectra 
yet there have been significant upgrades to the spectrophotometry with the latest reductions. 
Therefore, it is worth to revisit the line flux uncertainties. 

\begin{figure*}
\epsscale{.95} \plotone{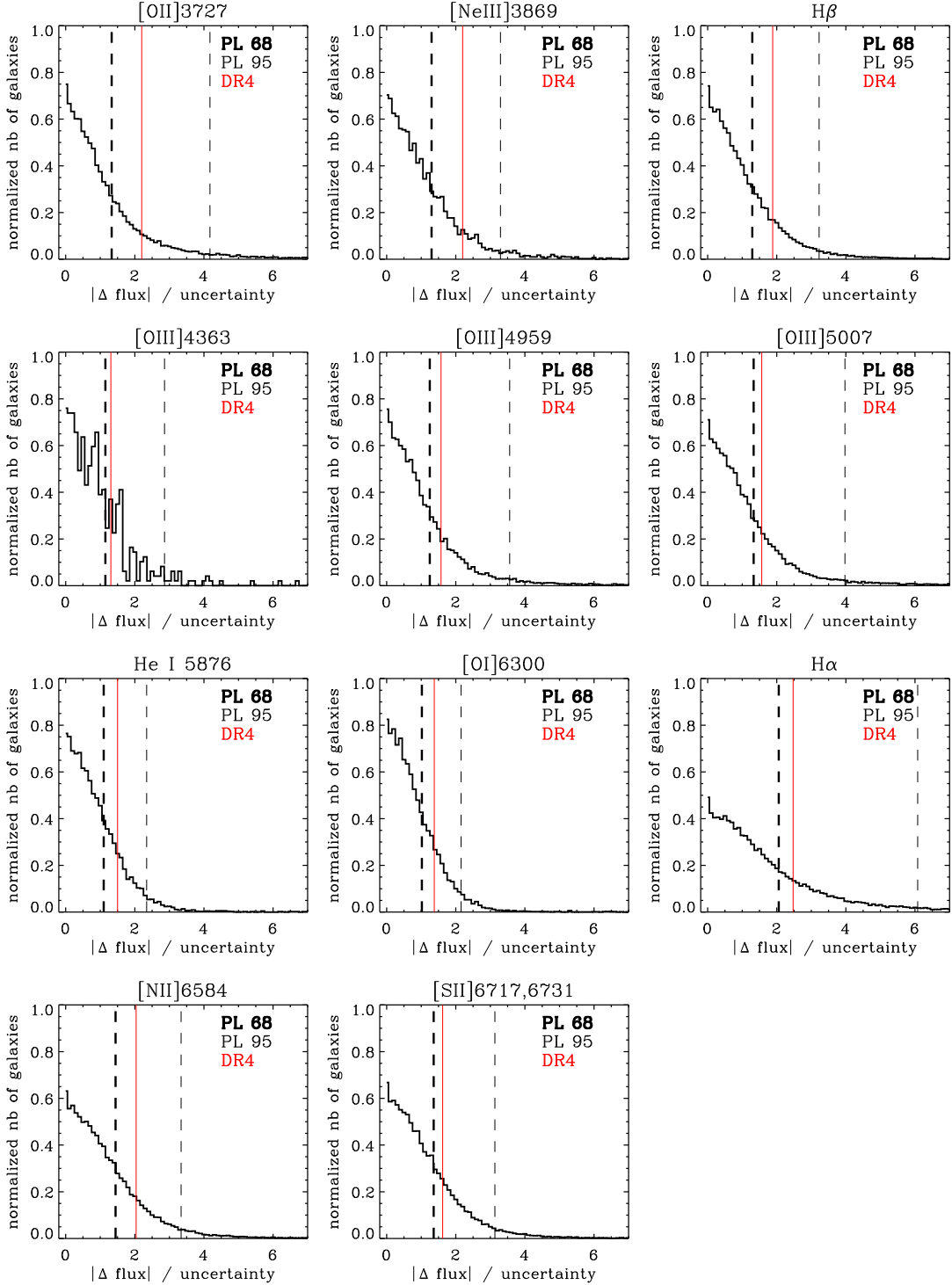}
\caption{
   Distribution of the absolute difference between flux measurements
   of duplicate observations of galaxies, normalized by the catalog flux
   uncertainties.  Each panel corresponds to a different emission
   line as labeled.  The 68th and 95th percentiles are respectively marked with
   thick and thin dashed lines, while the recommended uncertainty scaling factor of DR4 is
   indicated with the solid red line.
   }\label{fig:fluxUnc}
\end{figure*}

The MPA-JHU list of duplicate spectroscopic observations was used to compute the absolute 
value of the difference between multiple observations of the same targets.  Each pair of duplicata 
was considered. For a galaxy observed $N$ times, there are 
$\sum^{N-1}_{i=1}i$ different pairs. 
For each pair, we normalize the absolute difference by the uncertainty on the difference from the 
catalog (i.e., the two individual uncertainties added in quadrature).  If the catalog uncertainties 
corresponded to the true one-sigma uncertainties, 68\% of the cases would be within one.  Instead, 
the 68th percentile is always larger than one (Figure~\ref{fig:fluxUnc}).  

The results are tabulated for typical strong emission lines used in various galaxy evolution 
studies (Table~\ref{tab:fluxUnc}). Relative to the previous values found for DR4 measurements, 
these updated uncertainty are smaller, perhaps reflecting the improved spectrophotometry in the 
later reductions of the spectra. 

\begin{deluxetable}{ccc}
\tablecolumns{3}
\tablecaption{Scaling Factors for Line Flux Uncertainties}
\tablehead{
   \colhead{Emission Line}  &
   \colhead{DR7 Scale Factor} &
   \colhead{DR4 Scale Factor} }
\startdata
\oiilam\           &  1.33  & 2.199 \\
\neiiilam\         &  1.30  & 1.731 \\
\hb\               &  1.29  & 1.882 \\
\oiii$\lambda$4363 &  1.15  & 1.306 \\
\oiii$\lambda$4959 &  1.25  & 1.573 \\
\oiii$\lambda$5007 &  1.33  & 1.566 \\
\hei$\lambda$5876  &  1.10  & 1.501 \\
\oi$\lambda$6300   &  1.02  & 1.378 \\
\ha\               &  2.06  & 2.473 \\
\niilam\           &  1.44  & 2.039 \\
\siilam\           &  1.36  & 1.621 
\enddata
\label{tab:fluxUnc}
\end{deluxetable}

\begin{figure*}
\epsscale{1.} \plotone{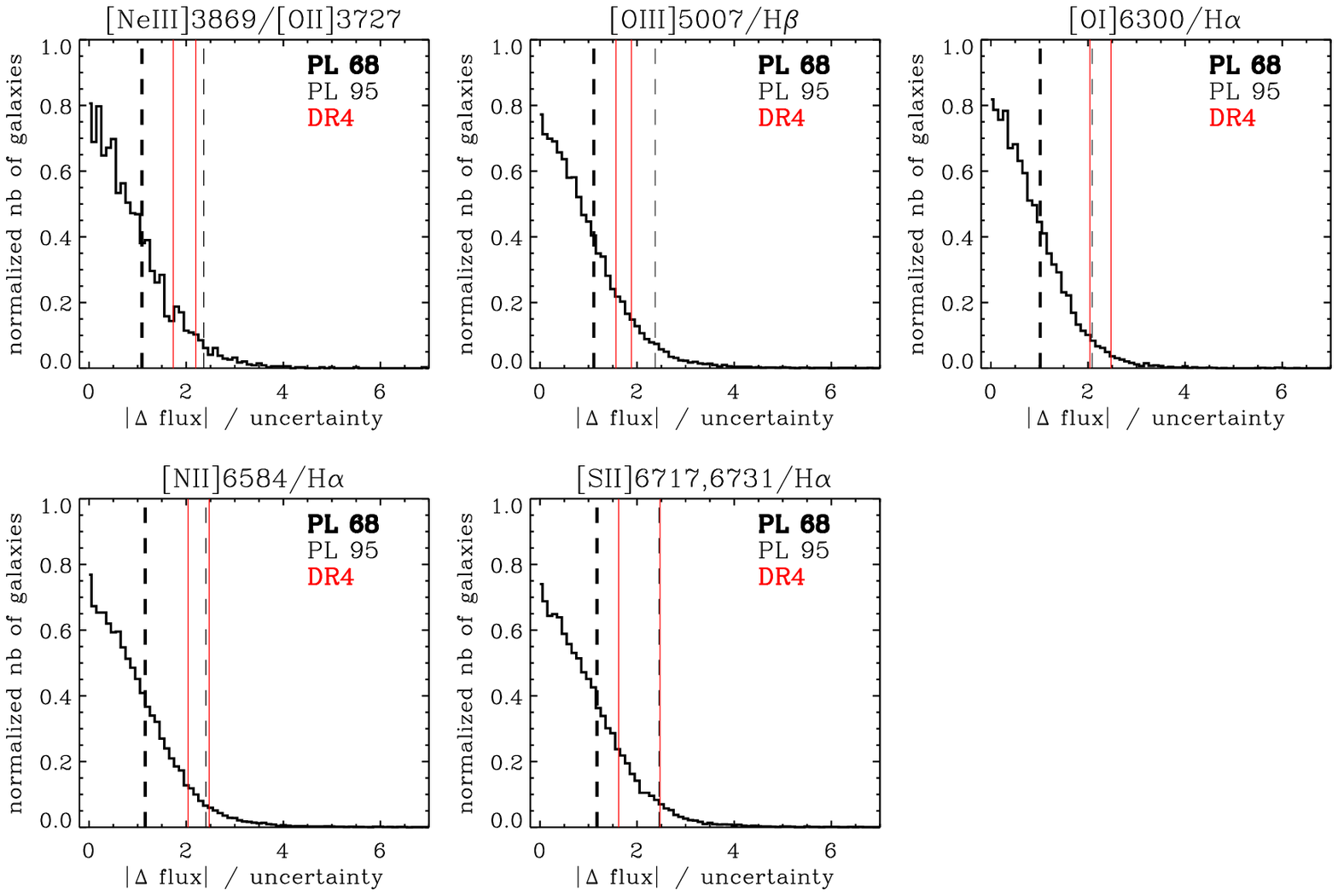}
\caption{
   Same as Figure~\ref{fig:fluxUnc} but for emission line flux ratios, as labelled.
   In each panel, the recommended uncertainty scaling factors of DR4 are 
   marked for the {\it individual} emission lines involved in the
   ratio (solid red lines; see Table~\ref{tab:fluxUnc}).
   }\label{fig:ratioUnc}
\end{figure*}

It is possible that some of this {\it true} albeit stasticial uncertainty may include 
multiplicative uncertainties that cancel out when taking the ratio of emission line fluxes.  
This may be especially relevant for ratios of lines that have similar wavelengths, in case 
of wavelength dependent uncertainties.  Thus, we perform the same calculation on commonly-used 
line ratios as for the individual line fluxes.  The results are displayed in Figure~\ref{fig:ratioUnc}, 
and listed in Table~\ref{tab:ratioUnc}.  The uncertainty scaling factors are generally 
smaller for the line ratios than for the individual line fluxes.  This trend supports the 
possibility that the individual line uncertainties include a multiplicative component that 
cancels out when taking the ratio of two lines.  In other words, line ratios are better reproduced 
between different spectral observations of the same galaxies than the absolute flux calibration.  
While not surprising, it is useful to quantify this effect for studies that depend on the 
sample selection and for a better understanding of the latter.

\begin{deluxetable}{ccc}
\tablecolumns{3}
\tablecaption{Scaling Factors for Uncertainties on Line Flux Ratios}
\tablehead{
  \colhead{Name} & \colhead{Emission Line Ratio} & \colhead{DR7 Scale Factor}}
\startdata
$Ne_3O_2$ &  \neiiilam/\oiilamboth\ &  1.09   \\
$O_3$     &  \oiiilam/\hb\          &  1.11   \\
\oi/\ha\  &  \oilam/\ha\            &  1.02   \\
\nii/\ha\ &  \niilam/\ha\           &  1.16   \\
\sii/\ha\ &  \siilam/\ha\           &  1.17   \\
$O_{32}$  &  \oiiilam/\oiilamboth\   &  1.25   \\
$R_2$     &  \oiilamboth/\hb\        &  1.23  \\
$N_2$     &  \niilamboth/\hb\        &  1.25  \\
$R_3$     &  \oiiilamboth/\hb\       &  1.11  \\
$R_{23}$   &  (\oiilam\ + \oiiilamboth)/\hb\  &  1.20  \\
\enddata
\label{tab:ratioUnc}
\end{deluxetable}

\section{B. Calculating and Modeling the Offsets on the MEx diagram}\label{sec:offset}

\begin{figure*}
\epsscale{0.9} \plotone{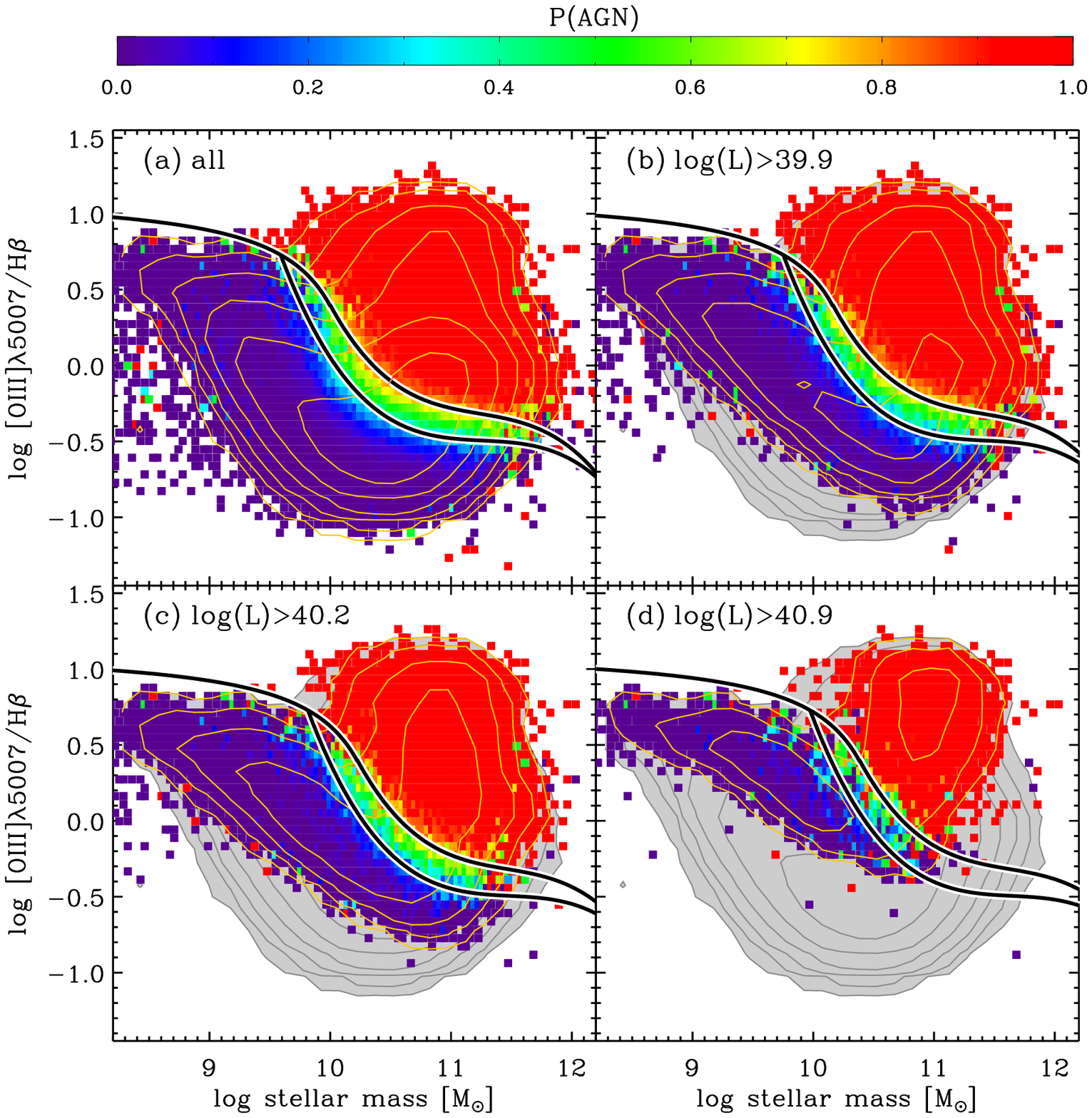}
\caption{
   MEx diagnostic diagram. 
   The dividing lines indicate regions corresponding to
   star-forming galaxies (below), MEx-intermediate galaxies (between)
   and MEx-AGN (above).  Colored points with thin contours show
   the distributions of $z\sim0$ SDSS for varying line luminosity
   threshold (as labeled), while the gray shaded contours include the
   full $z\sim0$ prior sample.  The color scheme indicates the AGN probability 
   based on the fraction of galaxies classified as BPT-AGN in each bin (color bar).
   The demarcation lines are offset between each panel to encompass the transition 
   region, where P(AGN) changes sharply from $\sim0.3$ to $\sim0.8$.  Contours are 
   logarithmically spaced (0.5~dex) with the outermost contour corresponding to ten 
   galaxies per bin of 0.15\,dex$\times$0.15\,dex.
}\label{fig:mexPagn}
\end{figure*}

The shape of the relations describing the MEx demarcation lines are held fixed according to 
equations~\ref{eq:up} and \ref{eq:lo}.  
We add an offset along the stellar mass axis ($x$) as a free parameter and, similarly to the 
appoach of Section~\ref{sec:revised}, we adjust the best-fit to go through the region of the MEx with 
$0.6 < P(AGN) < 0.85$ for the upper MEx curve. 
Figure~\ref{fig:mexPagn} illustrates the AGN probability on the MEx plane and the 
resulting fits for varying line luminosity thresolds.  

\begin{figure}
\epsscale{0.5} \plotone{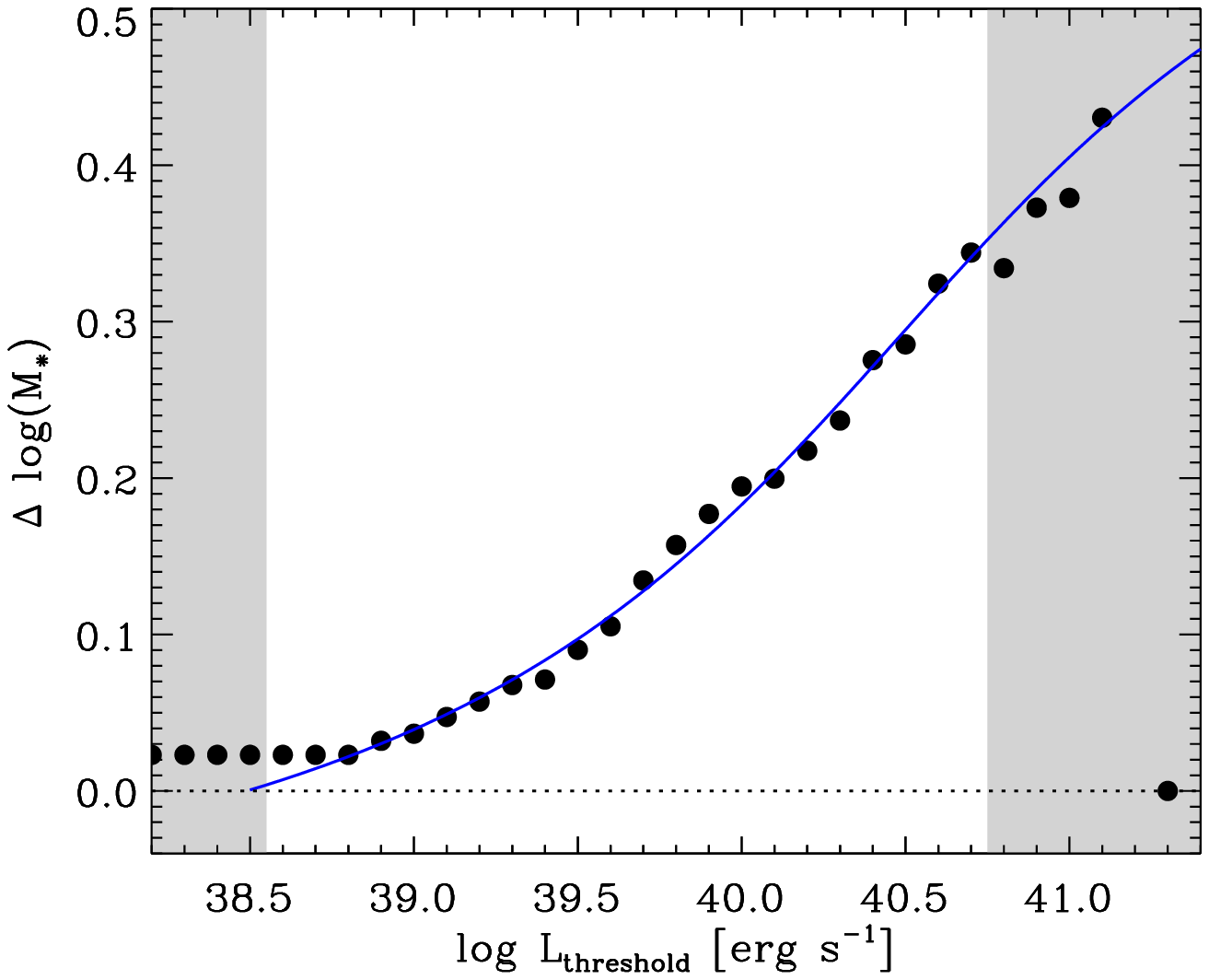}
\caption{
   Offsets along the stellar mass axis as a function of the line luminosity 
   threshold applied to \ha\ and \oiii.  
   The offsets were calculated separately for the upper (filled circles) 
   MEx curve and the spacing between the two curves is kept fixed.
   The data points were fitted with a simple analytical form over the range of 
   validity (excluding the gray zones; equation~\ref{eq:offset}).
}\label{fig:offset}
\end{figure}

We perform this exercise for line luminosity thresholds varying in steps of 0.1~dex and we compile 
the offsets in Figure~\ref{fig:offset}.  
The automated fitting is less constrained at high luminosities ($\log(L_{line}[\ergs])>40.7$) 
due to the small size of the prior sample when imposing a very high luminosity cut, and fails 
entirely above $\log(L_{line}[\ergs])>41.1$.  
Over the range of validity, we fit the offset in $\log(M_{\star})$ as a function of the 
threshold line luminosity with this functional form:
\begin{equation}\label{eq:offset}
   \Delta \log(M_{\star}) = a_0 + a_1 \times {\rm tan}^{-1}((\log(L_{line})-a_2)\times a_3),
\end{equation}
where $L_{line}$ is the threshold luminosity imposed for detection of \oiii\ and \ha.
The coefficients are \{0.28988, 0.28256, 40.479, 0.82960\}.

The offsets calculated with equation~\ref{eq:offset} allows one to visualize the regions 
of the MEx diagram largely populated by AGN or star-forming galaxies given an effective 
detection limit on \oiii\ and \ha\ (e.g., Figure~\ref{fig:mexPagn}).  More accurate computations of AGN 
(or star-forming) sample should take into account the full bivariate distribution of the prior sample.  
We developed routines\footnote{Availabe at https://sites.google.com/site/agndiagnostics/home/mex} 
to calculate the probability of having a given excitation type (star-forming, composite, 
LINER, Seyfert) given the location on the MEx diagram and the flux detection limit of the 
survey considered.  The calculations are made on a galaxy per galaxy basis, taking in each case 
the luminosity detection threshold at the corresponding redshift. Optionally, prescriptions 
for pure luminosity evolution of emission-line galaxies can be taken into account.  These 
prescriptions are described and applied in Section~\ref{sec:hiz} and in Appendix~\ref{app:altern}.

\section{C. Single-Line Selections}\label{app:altern}

\begin{figure}
\epsscale{0.9} \plotone{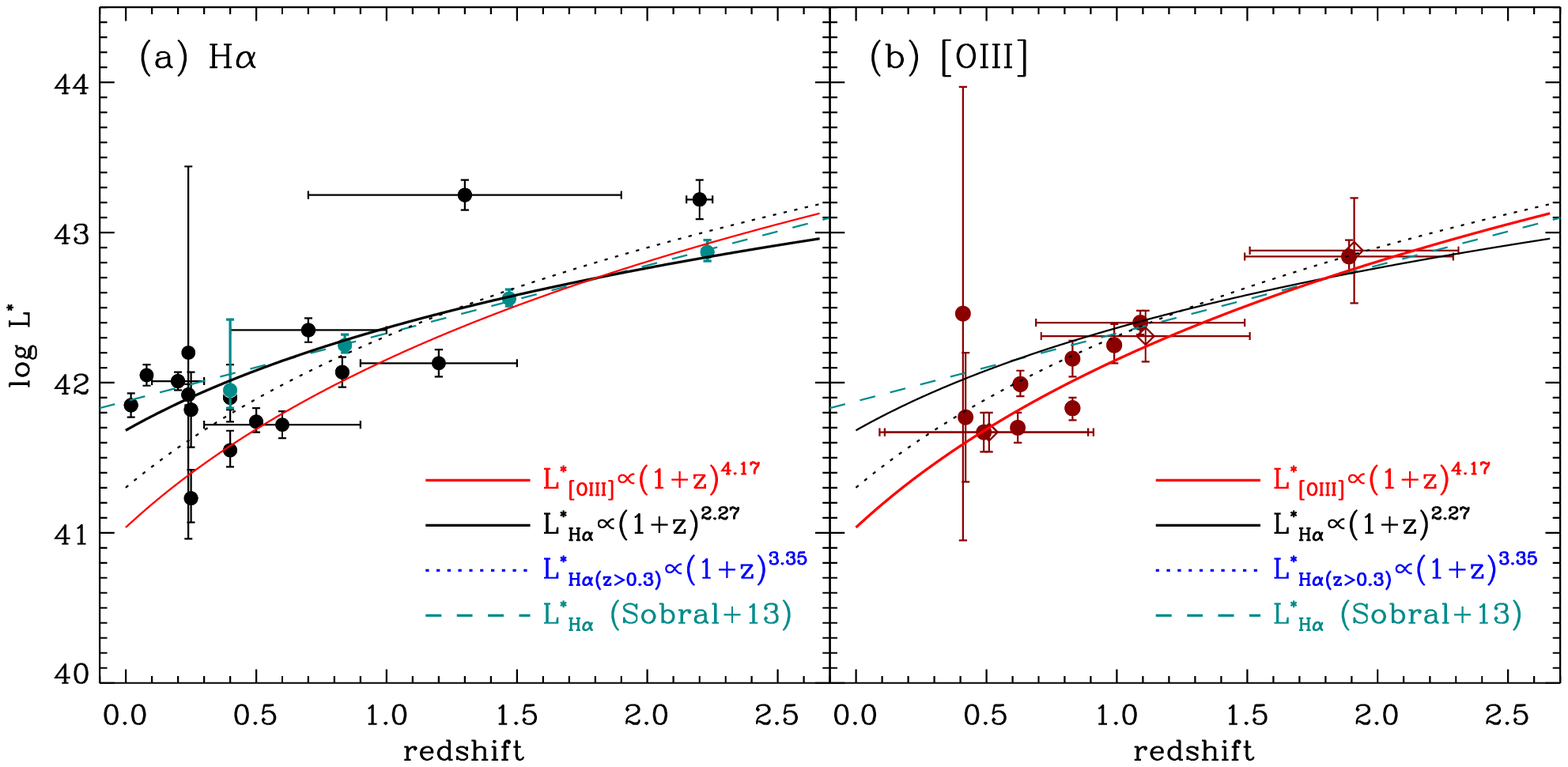}
\caption{
   Redshift evolution of the critical value $L^*$ of emission line luminosity 
   functions of (a) \ha, and (b) \oiii.  
   Values are compiled from the literature, and include a direct comparison with 
   the evolution found by \citet{sob13} in green (green filled circles and dashed line).
   Our best-fit evolution goes like (1+$z$)$^{2.27}$ for \ha\ (solid black curve on panel a) 
   and like (1+$z$)$^{4.17}$ for \oiii\ (solid red curve on panel b). We also fit to 
   \ha\ at $z>0.3$ (dotted line) to enable a closer comparison with the \oiii\ results.  
   In the case of \oiii, two open diamond symbols show repeated measurements  
   with fixed faint-end slope.  Using these two points instead of their counterpart at $z=1.1$ 
   and 1.9 (slightly offset for clarity of the plotting symbols) does not noticeably change the fit.
}\label{fig:Lstar}
\end{figure}

Alternatively to the scenario used in this work, one could consider a single emission line selection.  
We show a few such cases here, an \oiii\ selection where we substitute for the evolution of the knee 
of the \oiii\ luminosity function (Figure~\ref{fig:Lstar}b), an \ha-only selection based on the detection 
limit and evolution of the knee of the \ha\ luminosity function (Figure~\ref{fig:Lstar}a), an \hb-selection 
based on the detection limit of \hb\ but assuming the same evolution as the \ha\ luminosity function, and 
lastly, an \ha\ selection with luminosities taken at face values, without any $L^*$ evolution correction.  
We also consider this null evolution scenario with a two line selection requiring both \ha\ and \oiii\ to be 
more luminous than the detection luminosities.

To calculate the evolution of $L^*$ separately for \ha\ and \oiii\ emission lines, we compile values reported 
in the literature \citep{ly07,sob09,pir13,sob13,dra13,col13}, and apply a fit of the form 
$L^*(z) = L^*_{z=0} \times (1+z)^n$. We solve for $L^*$ at $z=0$ and index $n$ using routines from the 
MPFIT IDL package \citep{mar09}.

As shown in Figure~\ref{fig:Lstar}(a), the \ha\ evolution indicates a global fading of the galaxy population 
with time, which can be qualitatively understood as a consequence of decreasing SFRs in galaxies modulo 
dust attenuation, which is not included here.  We find that $\log(L^*_{\ha}) \propto$(1+$z$)$^{2.27}$ 
\citep[or a log-linear fit gives $\log(L^*_{\ha}) \propto 0.54z$, slightly steeper than the evolution 
of 0.45$z$ reported by][see green points and dashed line]{sob13}.  On the other hand, our overall best-fit 
evolution is slightly less steep than the reported evolution of sSFR for massive galaxies 
\citep[$\propto (1+z)^{2.8}$ for $10^{10}~M_{\sun}$][not shown]{sar13}, and than the $L^*_{\ha}$ fit restricted 
to $z>0.3$ (dotted line on Figure~\ref{fig:Lstar}a), i.e., the same redshift range over which \oiii\ can be fitted.  
Some of the variation between different studies of $L^*$ at the same redshifts could be due to cosmic variance 
or residual incompleteness in the sampling, but a detailed analysis is beyond the scope of this paper.  
Here, we adopt the best fit to all points for \ha\ (solid black line), which is similar to the results from \citet{sob13}.

For \oiiilam, the best-fit evolution is steeper than for \ha.  
We find $\log(L^*_{\oiii}) \propto$(1+$z$)$^{4.17}$ (red curve in Figure~\ref{fig:Lstar}), while a log-linear 
fit with redshift yields $\log(L^*_{\oiii}) \propto 0.85z$.  However, there is a scarcity of measurements 
of galaxy \oiii\ LFs at the lowest redshifts.  Namely, we did not find such measurements at $z<0.3$.  
If similar to the \ha\ situation, we could expect substantial scatter at a fixed redshift including at the 
lowest redshifts, implying that the true evolution of $L^*_{\oiii}$ could be different from the best-fit 
shown here and more similar to the \ha\ evolution.  Indeed, fitting only $z>0.3$ points for \ha\ yields a 
steeper evolution, though still not as steep as our current best-fit case for \oiii.  This reinforces the 
relevance to try two evolutionary scenarios for \oiii\ by applying the \ha\ evolution (panel a) and the 
steeper \oiii\ evolution (panel b), with the assumption that these two scenarios may bracket the real 
evolution.

Physically, we could expect that the steeper evolution seen for \oiii\ could be a real feature because 
it is qualitatively consistent with our understanding of gas-phase metallicity evolution in the sense 
that higher redshift galaxies had lower metallicities therefore comparatively more luminous \oiii\ lines. 
Thus, the \oiii\ luminosity function fading with cosmic time would then be a combination of 
both decreasing SFRs and increasing gas-phase metallicities in the bulk of star-forming galaxies.

\subsection{\oiii\ Selection}

We investigate a pure \oiii\ selection by applying luminosity 
thresholds to \oiii\ corresponding to the detection limit to which we subtract the 
evolution of $L^*_{\oiii}$ shown in Figure~\ref{fig:Lstar}(b) between the redshift of the sample 
considered and the prior reference sample at $z\sim0.09$.

\begin{figure*}
\epsscale{.95} \plotone{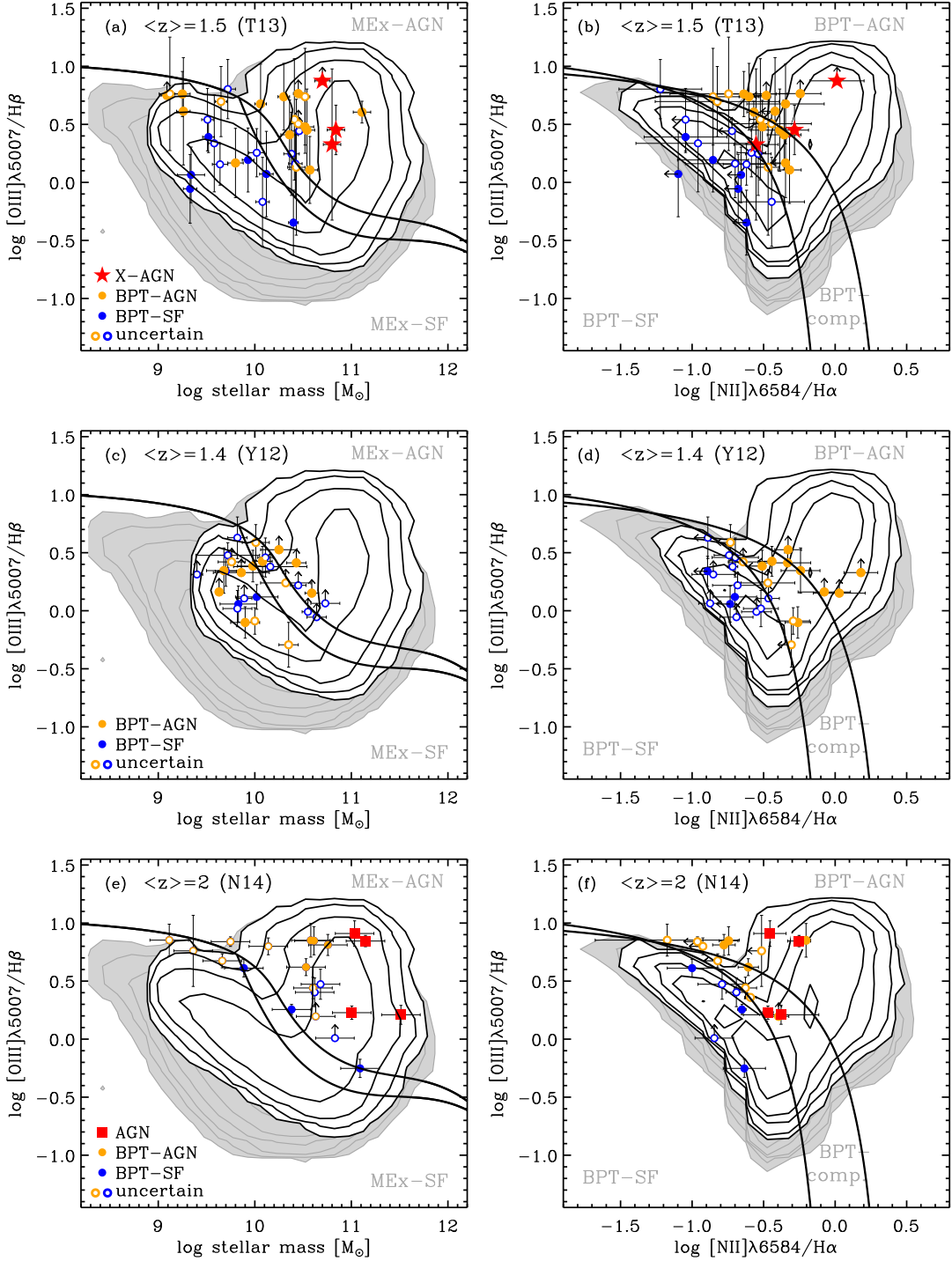}
\caption{
  Same as Figure~\ref{fig:hizbpt} with predicted contours corresponding to a pure \oiii\ selection, 
  after accounting for $L^*_{\oiii}$ evolution. 
}\label{fig:hizbptOiii}
\end{figure*}

In Figure~\ref{fig:hizbptOiii}, we repeat the comparison with the same high-redshift samples as 
in Figure~\ref{fig:hizbpt}. The figures are identical except for the predicted contours, which 
now correspond to pure \oiii\ selection and $L^*_{\oiii}$ evolution.  The main differences are 
that (i) the AGN branch is broader and extends both to the left-most reach of the prior sample, which was 
previously removed by the \ha\ luminosity cut, and further to the right toward the locus of LINERs, 
which was previously removed with a more luminous \oiii\ cut, and (ii) the star-forming branch reaches to lower 
values of \oiii/\hb, which was previously not the case due to requiring \oiii\ to be as luminous 
as \ha\ and applying the higher threshold corresponding to milder $L^*_{\ha}$ evolution relative 
to $L^*_{\oiii}$ evolution. This extension of the star-forming branch to lower \oiii/\hb\ values 
worsens the agreement especially at the massive, high \nii/\ha\ end of the diagrams. 
Similarly, the extension of the AGN contours toward the LINERs (higher stellar masses on the MEx, 
and higher \nii/\ha\ on the BPT on the AGN side) worsens the agreement on the BPT for the T13 and N14 samples, 
and is unconstrained for the Y12 sample as the authors had previously removed independently identified 
AGNs from their parent sample and are left with fewer AGN candidates, several of them having only lower 
limits on \oiii/\hb\ ratios.  However, the agreement on the AGN side of the MEx is slightly improved for the 
N14 sample, which tends to be biased toward massive hosts compared to the $z\sim1.5$ samples.

\begin{figure*}
\epsscale{1.} \plotone{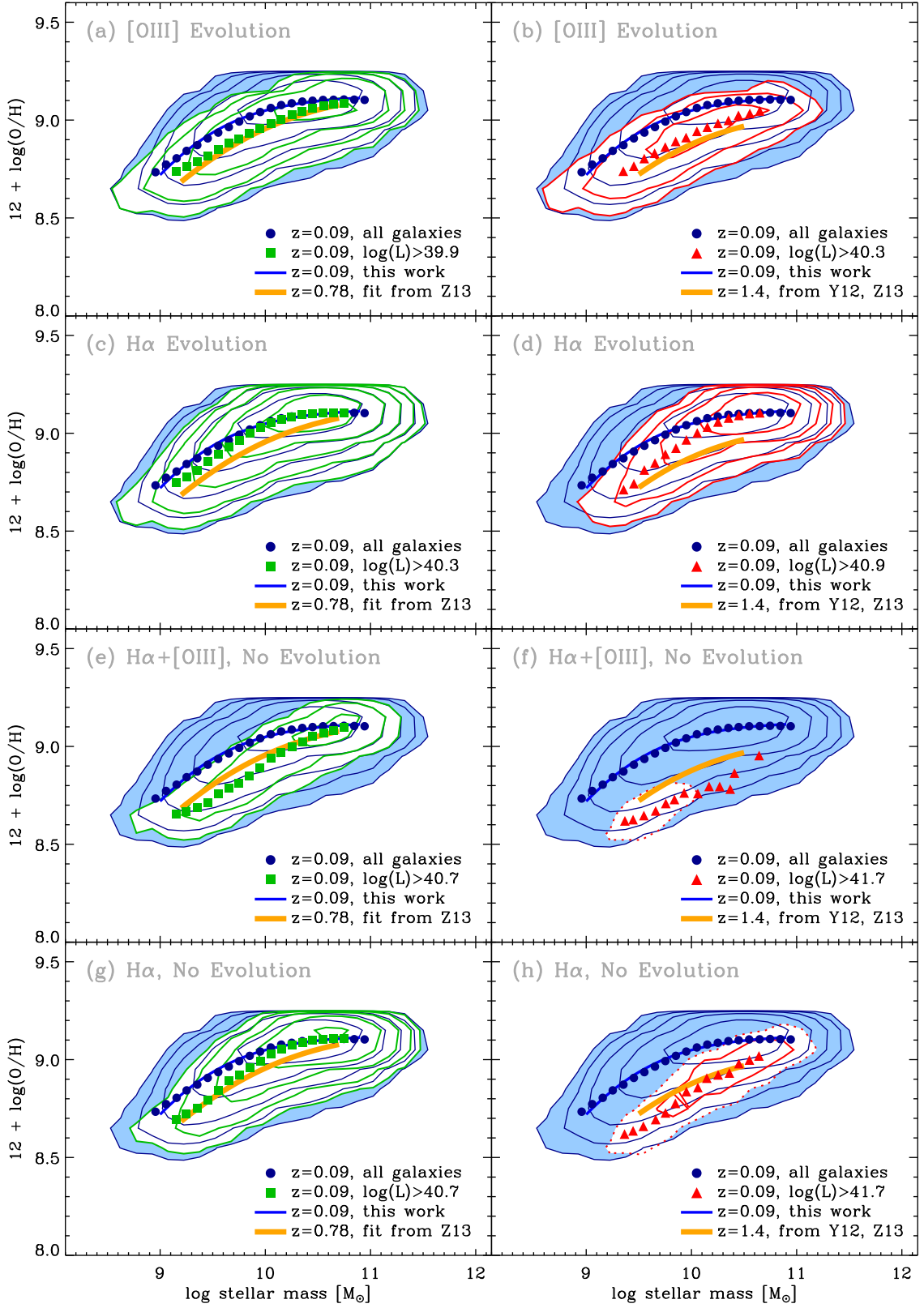}
\caption{
   MZ relation. Each row contains a pair of panels corresponding to the labeled scenario.  The left-hand 
   panel includes a comparison to observed $z\sim0.8$ relation from Z13, and the right-hand panel 
   includes a comparison to observed $z\sim1.4$ relation from the Y12 sample fitted by Z13 (orange line).  
   The meaning of plotting symbols and contours is identical to Figure~\ref{fig:MZ}, except that the 
   threshold line luminosities to make the $z\sim0.8$ (green) and $z\sim1.4$ (red) SDSS predictions 
   are adjusted individually according to selection and evolution scenarios, as follows: 
     (a,b) \oiii\ selection and evolution of $L^*_{\oiii}$;
     (c,d) \ha\ selection and evolution of $L^*_{\ha}$; 
     (e,f) \ha+\oiii\ selection and no evolution; 
     (g,h) \ha\ selection no evolution. 
   Contours are logarithmically spaced (0.5~dex) with the outermost contour corresponding to 50 
   galaxies per bin of 0.15\,dex$\times$0.15\,dex.  In panels (f) and (h), we show one additional 
   outer contour at 0.5~dex lower number density (dotted red contour) to aid the visualization of those 
   less populated subsamples.
}\label{fig:MZOiii}
\end{figure*}

By fading the \oiii\ luminosities more strongly than in our fiducial approach, we draw a 
comparison sample that may include more metal-rich galaxies in the prior sample relative to the 
target higher-redshift sample.  
To compare with the main method adopted in this work, we revisit the predicted $MZ$ relations 
in Figure~\ref{fig:MZOiii}(a,b). Relative to $MZ$ relations predicted with the fiducial approach 
(Figure~\ref{fig:MZ}), the new predicted contours for $z\sim0.8$ and $z\sim1.4$ samples indeed include 
galaxies that are slightly more metal-rich, and more noticeably at higher masses ($>10^{10}$~\Msun) 
in the $z\sim1.4$ Y12 sample.  Qualitatively, the predicted trends go in the same direction as 
the observations, and are in fair agreement with the observations.  Compared with our fiducial 
scenario from Figure~\ref{fig:MZ}, the agreement may be slightly better for the $z\sim0.8$ sample, and 
slightly worse for the $z\sim1.4$ sample, but not strikingly different.

\subsection{\hb\ Selection}

\begin{figure*}
\epsscale{.95} \plotone{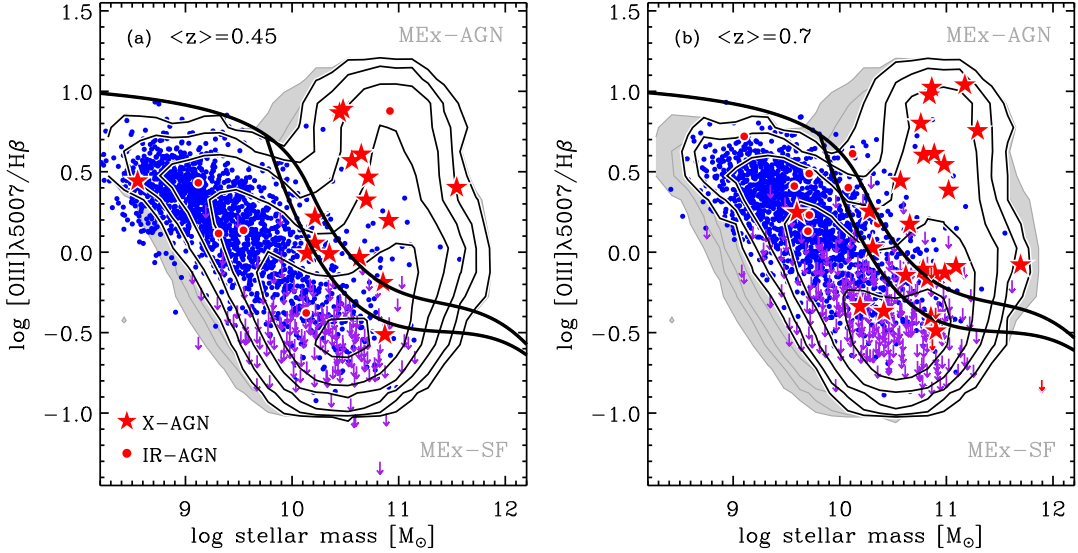}
\caption{
   MEx AGN diagnostic diagram applied to intermediate redshift galaxies:   
   (a) $\langle z \rangle=0.45$, 
   (b)  $\langle z \rangle=0.7$.
   Same as Figure~\ref{fig:hiz} except for an \hb\ selected sample.  The purple 
   downward arrows indicate \oiii\ upper limits and \hb\ detections, and black 
   contours show the corresponding \hb\ selected prior sample. Contours are 
   logarithmically spaced (0.5~dex) with the outermost contour corresponding to ten 
   galaxies per bin of 0.15\,dex$\times$0.15\,dex. 
}\label{fig:hizHb}
\end{figure*}

At intermediate redshifts where we only have \hb\ and \oiii, and no coverage of \ha, we  
perform a pure \hb\ selection by keeping only galaxies with \hb\ 
detections (3$\sigma$) and including cases where \oiii\ is an upper limit.  We use the 
same $z\sim0.45$ and $z\sim0.7$ samples from J11 as in Section~\ref{sec:hizsamples} except 
that we now remove \hb\ upper limits, and add instead points with \hb\ detections and 
\oiii\ upper limits (purple arrows on Figure~\ref{fig:hizHb}).
The predicted contours include prior galaxies with \hb\ luminosities above the line flux 
detection threshold faded by the same evolution as \ha, and reproduce fairly well the 
locus of the observed points, though the constraint is weaker at low \oiii/\hb\ ratios 
because most observations have only an \oiii\ upper limit in these cases.

Relative to the full prior sample (filled gray contours), this \hb\ selection cuts 
the left-hand side of both the star-forming and AGN branches.  While an \oiii\ selection 
formally cuts the bottom part of the branches.  The results are fairly similar on the star-forming 
side because of the tilt of the branch from top left to bottom right.  In details, the selections 
differ most strongly at the high-mass end of the star-forming branch, which is more trimmed with 
an \oiii\ selection, and at the low-mass end of the AGN branch, which is more trimmed for a \hb\ 
selection.

\subsection{\ha\ Selection}

\begin{figure*}
\epsscale{0.95} \plotone{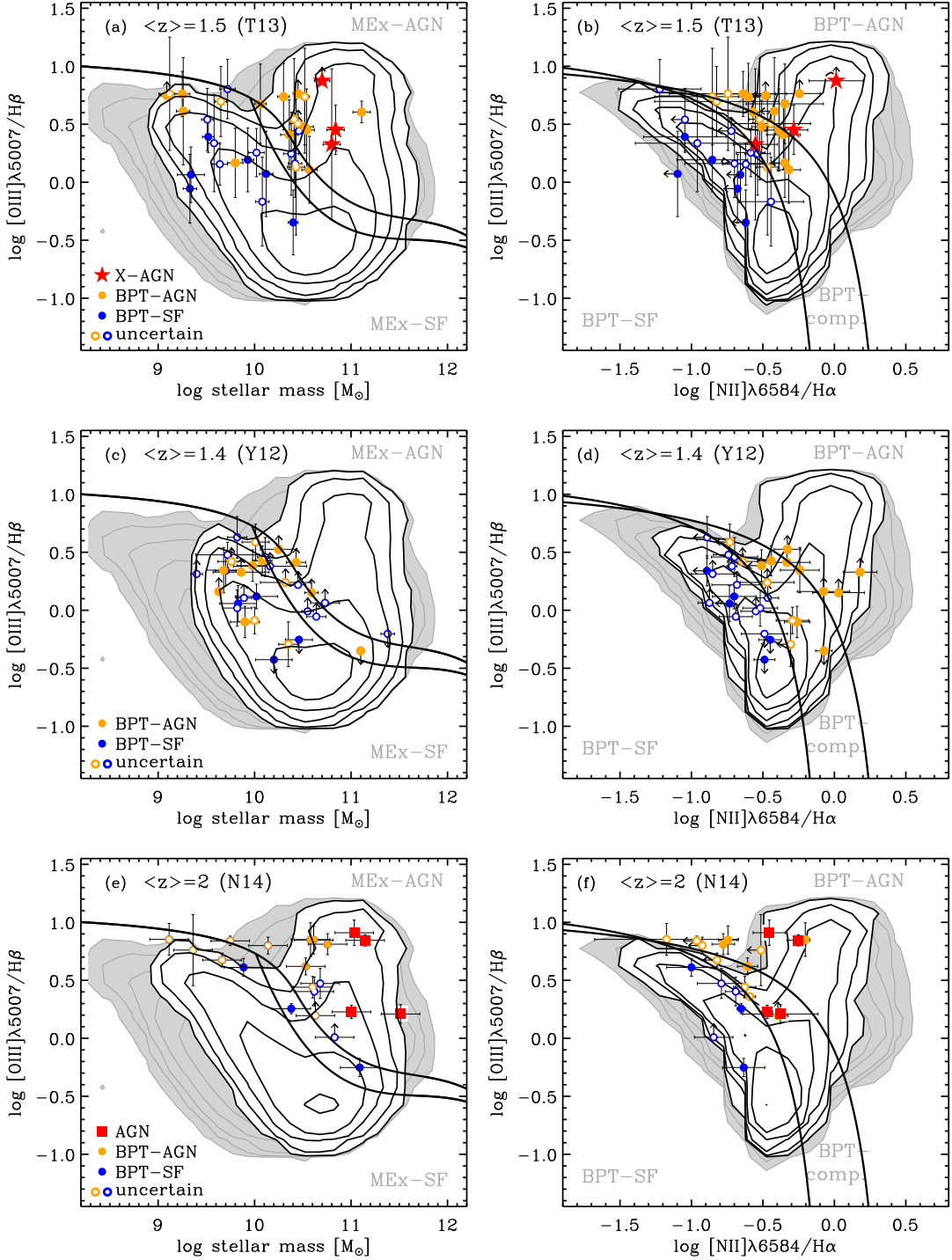}
\caption{
  Same as Figure~\ref{fig:hizbpt} with predicted contours corresponding to a pure \ha\ selection, 
  and including $L^*_{\ha}$ evolution. 
}\label{fig:hizbptHa}
\end{figure*}

We repeat the analysis with a pure \ha\ selection for the high-redshift samples at $z>1$. 
The new predicted contours are shown in Figure~\ref{fig:hizbptHa}. The observed samples are 
the same as in Figure~\ref{fig:hizbpt} except that we now include galaxies which have an 
upper limit on \oiii\ and therefore on their \oiii/\hb\ ratios (only present in the Y12 sample). 
The \ha\ selection contours appear to overpredict the number 
of galaxies that should be observed at low \oiii/\hb\ ratios.  We recall that the T13 sample 
was selected from HST/grism data covering \hb\ and \oiii, and is effectively \oiii-selected. 
While in principle \hb\ could be brighter than \oiii, this would require very elevated  
SFRs given the bright flux limit ($5\times10^{-17}$~\ergs\,cm$^{-2}$) and that \hb\ is intrinsically 
much fainter than \ha\ (\hb/\ha\ $<$ 0.35) while at the same time \oiii/\ha\ is often higher at 
high emission line luminosities \citep{ly07,col13}. 

The Y12 parent selection includes as a criterion predicted \ha\ fluxes, but the subsample shown here 
is not purely \ha-selected because some galaxies with \ha\ detection could have upper 
limits to both \oiii\ and \hb, preventing us from constraining their position on the BPT and 
MEx diagrams. Keeping this caveat in mind, Y12 still appear to be the sample best fitted by \ha\ 
selected contours on the BPT diagram relative to the T13 and N14 samples.  The N14 sample is 
poorly represented by the contours on the star-forming side of both the MEx and BPT diagrams, with a 
clear overprediction of the number of galaxies with low \oiii/\hb\ ratios.  The contours on 
the MEx-AGN region encompass up to 73\% (11/15) of galaxies including uncertainties, while the 
BPT contours appear to be a poorer representation of the data on the AGN branch by overpredicting 
cases with comparatively elevated \nii/\ha.

Like the \oiii\ line, \ha\ is also sensitive to SFRs but less to metallicities.  Thus, we expect that 
predicted $MZ$ relations based on \ha\ could reach higher metallicities than an \oiii\ selection 
with the same luminosity threshold. This is indeed what we find in Figure~\ref{fig:MZOiii}(c,d).  
Galaxies with the highest \ha\ luminosities still tend to have lower metallicities at the low-mass end, but 
this is no longer the case at the high-mass end, where they have comparable metallicities to the 
full prior sample.  This is particularly visible in panel (d), where the predicted red contours and triangles  
correspond to \ha\ luminosity above $10^{40.9}$~\ergs without constraints to other emission lines.

Overall, a pure \ha\ selection does not seem appropriate for some of the samples shown in 
Figures~\ref{fig:hizbptHa} and \ref{fig:MZOiii}.  However, this exercise demonstrates that a 
an \ha\ selection should in principle yield more metal-rich galaxies at high masses and 
therefore a weaker apparent $MZ$ evolution at the highest mass relative to a sample for which 
it is required to detect other emission lines such as \oiii. Also, comparing the two scenarios from 
Figure~\ref{fig:MZOiii}, we can predict that \ha\ selected samples may thus find a steeper $MZ$ 
relation than an \oiii-selected sample, or a sample requiring both \oiii\ and \ha.   
This is at least qualitatively consistent with recent work by \citep{zah14}.  These authors 
find a steepening of the $MZ$ relation for their $z\sim1.6$ sample of \ha-selected galaxies 
\citep[parent sample described by][]{kas13}.

\subsection{Null Evolution Scenario}

Lastly, we consider the case where the luminosity detection threshold is taken at face value, i.e., 
without applying any corrections for the fading of emission line $L^*$.  In one case, we require both 
\ha\ and \oiii\ to be more luminous than the detection limit at the median redshift of the sample 
(Figure~\ref{fig:MZOiii}e,f).  In the other case, the requirement is only set on \ha\ 
(Figure~\ref{fig:MZOiii}g,h).  The resulting $MZ$ predictions are made for the 
same $z\sim0.8$ and $z\sim1.4$ samples considered with alternative scenarios. 
When requiring both \ha\ and \oiii\ to be above the luminosity detection threshold, the $MZ$ 
relation evolution is clearly overpredicted in both redshift slices, with 
the contours lying at lower metallicities than the observed relations.  This implies that 
comparing high-redshift galaxies to lower redshift galaxies of the same \ha\ and \oiii\ 
luminosities may be too extreme as this scenario overshoots the observed evolution.  
On the other hand, predictions from the null evolution scenario applied only to \ha\ lie 
closer to the observed $MZ$ relations at $z\sim0.8$ and $z\sim1.4$.  The slope of the predicted 
relation is slightly steeper than observed, but not inconsistent given typical uncertainties 
that characterize metallicity measurements, and the spread around the mean.  

For the two $MZ$ relation samples considered here, at $z\sim0.8$ and $z\sim1.4$, the empirical 
predictions that correspond the most closely to the observations are the fiducial scenario with 
both \ha\ and \oiii\ lines above the threshold faded by $L^*_{\ha}$ evolution, the \oiii\ 
only selection faded by $L^*_{\oiii}$ evolution, and the \ha-only selection with no evolution.  
The data in hand do not allow us to rule out any of these three options.  
However, the least prefered scenarios are a pure \ha\ selection with $L^*_{\ha}$ evolution, 
a null evolution scenario with both \ha\ and \oiii\ above the detection threshold, and using the 
full SDSS prior sample (blue contours and blue circles).  If we may not have singled out 
the best scenario in absolute terms, we have demonstrated that a few sensible options are 
preferable than comparing high-redshift samples to the full SDSS low-redshift sample with no 
account of selection effects arising from emission line detection limits.

\section{D. Application to Stacked Spectra}\label{app:stac}

\begin{figure}
\epsscale{0.45} \plotone{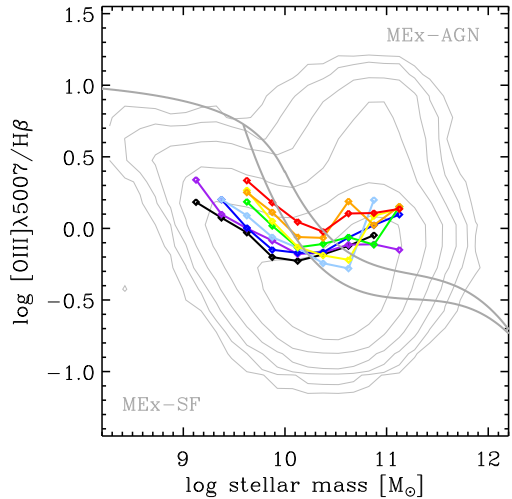}
\caption{
   MEx diagnostic diagram of the eight redshift slices used by \citet{vit13}.  The redshift slices 
   are centered at $z=$0.21 (black), 0.27 (purple), 0.34 (blue), 0.40 (light blue), 0.59 (green), 
   0.67 (yellow), 0.76 (orange) and 0.84 (red).  
   In each slice, spectra were stacked in bins of stellar mass, and the \oiii/\hb\ emission 
   line ratio was measured on the stacked spectra.  A zoomed in version of these tracks was 
   presented in Figure~6 of the article by \citet{vit13}.  Contours show the number density of 
   our reference $z\sim0$ SDSS sample, and are logarithmically spaced by 0.5~dex, with the 
   outermost contour corresponding to 10 galaxies per bin (0.15~dex$\times$0.15~dex).
}\label{fig:vitall}
\end{figure}

In this Appendix, we apply our empirical prediction method to observational results published 
by \citet{vit13}.  These authors stacked zCOSMOS-bright spectra in bins of redshift and stellar masses and 
found that the redshift tracks were systematically displaced from one other on the MEx diagram, 
in the sense that the higher redshift bins tend to be located above and/or to the right of the 
lower redshift cases (Figure~\ref{fig:vitall}). 
 
\begin{figure*}
\epsscale{1.1} \plotone{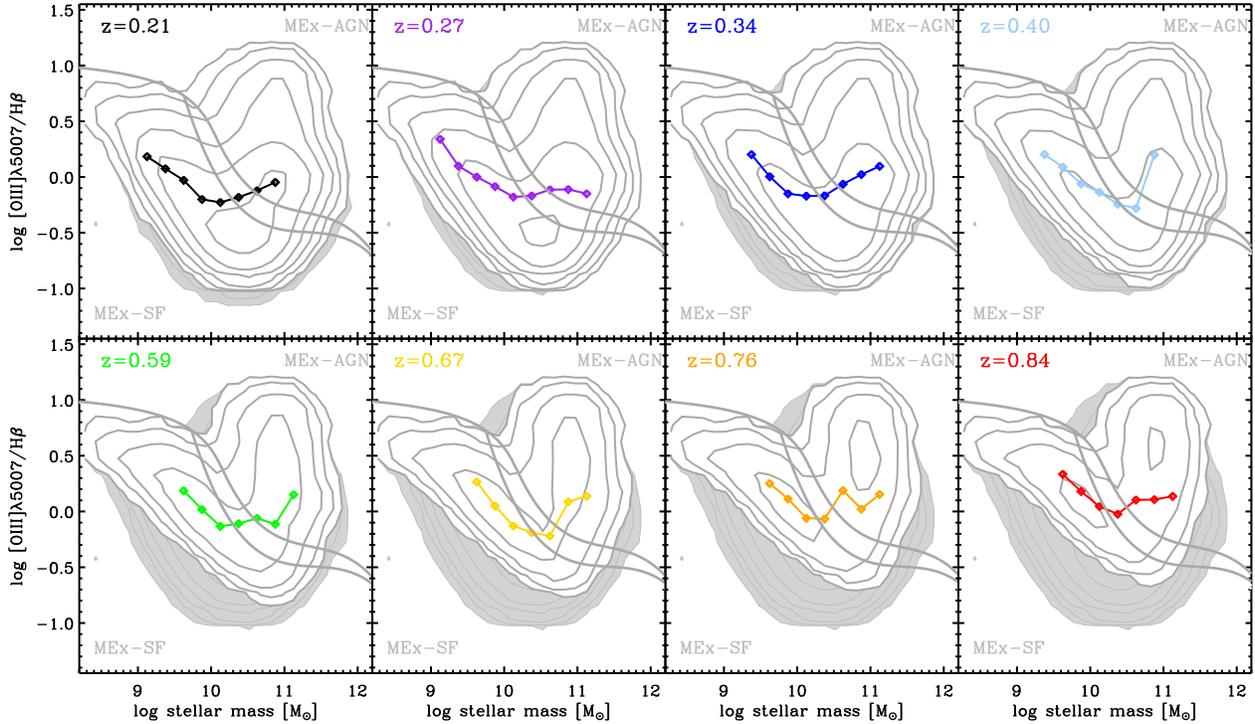}
\caption{
   MEx diagnostic diagrams showing stacked spectra results from \citet{vit13} in bins of stellar mass and redshift 
   (colored points).  
   Each panel contains one of the eight narrow redshift slices, labeled with the median redshift in the top left. 
   On a given panel, the individual points correspond to stacked values in bins of stellar mass, the underlying 
   filled gray contours encompass the full SDSS emission-line sample, while the gray contours illustrate the 
   prediction from the prior sample tailored to the central redshift of each bin, and incorporating selection 
   effect by applying the minimum line luminosity to \oiii\ and \ha\ in the prior sample.
   The tailored prior samples trace especially well the observed star-forming branch.
}\label{fig:vitale}
\end{figure*}

Here, we attempt to reproduce these trends using a fixed flux detection limit of $10^{-17}$~\ergs\,cm$^{-2}$ 
(M. Mignoli, 2013, private communication). 
In most redshift bins, the SDSS prior samples corresponding to the detection limit 
of the central redshift appear to provide us with a good representation of the star-forming branch 
(Figure~\ref{fig:vitale}) with an overall satisfactory qualitative agreement.  In details, the AGN side 
does not show a single behavior from one redshift bin to another, and the stacked spectra tend to 
have fairly flat \oiiihb\ ratios at the highest masses, while the SDSS prior distributions predict 
a rising AGN branch.  The cause for this difference is not known but it could be related to the stacked 
observations containing many more non-detections than detections and/or to greater uncertainties in 
the line ratios of stacked spectra at the highest masses.  The uncertainties are not included in the 
work by \citet{vit13} but the numbers of objects in the highest mass bins are somewhat smaller than in 
the intermediate mass bins (their Table~1).  

\bibliographystyle{apj_sj}
\bibliography{ms_selec_evol}

\end{document}